\documentclass[aps,prb,reprint,twocolumn, superscriptaddress]{revtex4-1}
\usepackage{bbold}
\usepackage{amsmath,amsfonts,amsmath,mathtools,bbm,bm}
\usepackage{graphicx}
\usepackage{braket}
\usepackage[colorlinks,linkcolor=blue,urlcolor=blue,citecolor=blue]{hyperref}
\usepackage[all]{hypcap}

\usepackage{blindtext}
\usepackage{enumitem}
\usepackage{xcolor}
\usepackage[normalem]{ulem}
\usepackage{float}
\setcitestyle{numbers,square}

\newcommand{\ii}{\mathrm{i}}

\newcommand{\Tr}{\mathrm{Tr}}
\newcommand{\ui}{\mathrm{i}}

\def\ii{\mathrm{i}}

\def\({\left (}
\def\){\right )}

\graphicspath{{Figure/PNG/}{Figure/PDF/}{Figure/EPS/}{Figure/TEX/}{Figure/}}

\newcommand\mydots{\hbox to 1em{.\hss.\hss.}}

\begin{document}

\title{Full Eigenstate Thermalization via Free Cumulants in Quantum Lattice Systems}

\author{Silvia Pappalardi}
\email{pappalardi@thp.uni-koeln.de}
\affiliation{Laboratoire de Physique de l’\'Ecole Normale Sup\'erieure, ENS,  CNRS, F-75005 Paris, France}
\affiliation{Institut f\"ur Theoretische Physik, Universit\"at zu K\"oln, Z\"ulpicher Straße 77, 50937 K\"oln, Germany}

\author{Felix Fritzsch}
\affiliation{Physics Department, Faculty of Mathematics and Physics, University of Ljubljana,  SI-1000, Slovenia}
\affiliation{Max Planck Institute for the Physics of Complex Systems, 01087 Dresden, Germany}

\author{Toma\v{z} Prosen}
\affiliation{Physics Department, Faculty of Mathematics and Physics, University of Ljubljana,  SI-1000, Slovenia}
\affiliation{Institute of Mathematics, Physics and Mechanics, Ljubljana, SI-1000, Slovenia}

\date{\today}

\begin{abstract}
The Eigenstate-Thermalization-Hypothesis (ETH) has been established as the general framework to understand quantum statistical mechanics.  Only recently has the attention been paid to so-called full ETH, which accounts for higher-order correlations among matrix elements, and that can be rationalized theoretically using the language of Free Probability.  
In this work, we perform the first numerical investigation of the full ETH in physical many-body systems with local interactions by testing the decomposition of higher-order correlators into thermal free cumulants for local operators. 
We perform exact diagonalization on two classes of local non-integrable (chaotic) quantum many-body systems: 
spin chain Hamiltonians and Floquet brickwork unitary circuits. We show that the dynamics of four-time correlation functions are encoded in fourth-order free cumulants, as predicted by ETH. Their dependence on frequency encodes the physical properties of local many-body systems and distinguishes them from structureless, rotationally invariant ensembles of random matrices.
\end{abstract}

\maketitle

{\bf{Introduction}} - Understanding the emergence of statistical mechanics in quantum many-body systems is a longstanding challenging problem \cite{gallavotti1999statistical, gutzwiller2013chaos}. The most established framework for this purpose is the Eigenstate Thermalization Hypothesis (ETH) \cite{deutsch1991quantum, srednicki1999approach, dalessio2016from}, which combines the random matrix nature of chaotic spectra \cite{casati1980connection, berry1981quantizing, bohigas1984characterization} with a focus on physical observables \cite{vonneumann1929beweis}.  According to ETH, local observables $\hat A$ in the energy eigenbasis are pseudorandom matrices, whose statistical properties are smooth thermodynamic functions. 
The validity of ETH has been well established by a myriad of numerical studies on non-integrable local many-body systems 
\cite{prosen1999,rigol2008thermalization, biroli2010effect,  polkovnikov2011colloquium, ikeda2013finite, steinigeweg2013eigenstate,alba2015eigenstate, beugeling2015off, luitz2016long, luitz2016anomalous, leblond2020eigenstate, brenes2020eigenstate, fritzsch2021eigenstate, garratt2021pairing}. 
 However, the standard ETH neglects correlations and it describes equilibrium up to two-point dynamical functions, leaving unanswered the question about \emph{multi-point correlators}. These are the relevant quantities beyond linear response,  to characterize higher-order hydrodynamics \cite{doyon2020fluctuations, myers2020transport, fava2021hydrodynamic}, quantum chaos and scrambling (via the out-of-time order correlators) \cite{maldacena2016bound, hosur2016chaos, roberts2017chaos, xu2022scrambling, garcia2022out} or for novel concepts such as deep thermalization \cite{cotler2023emergent,ho2022exact, claeys2022emergent, ippoliti2022solvable, lucas2022generalized}.  The importance of correlations among the matrix elements, already pointed out in the semi-classical limit in Ref.\cite{prosen1994statistical}, has lately attracted a lot of interest from the many-body \cite{foini2019eigenstate, foini2019eigenstate2, chan2019eigenstate, murthy2019bounds, richter2020eigenstate, wang2021eigenstate, brenes2021out,  dymarsky2022bound, nussinov2022exact} to the high-energy communities \cite{sonner2017eigenstate, jafferis2022matrix,jafferis2022jt}. \\
\begin{figure}[t]
	\centering
	\includegraphics[width=1 \linewidth]{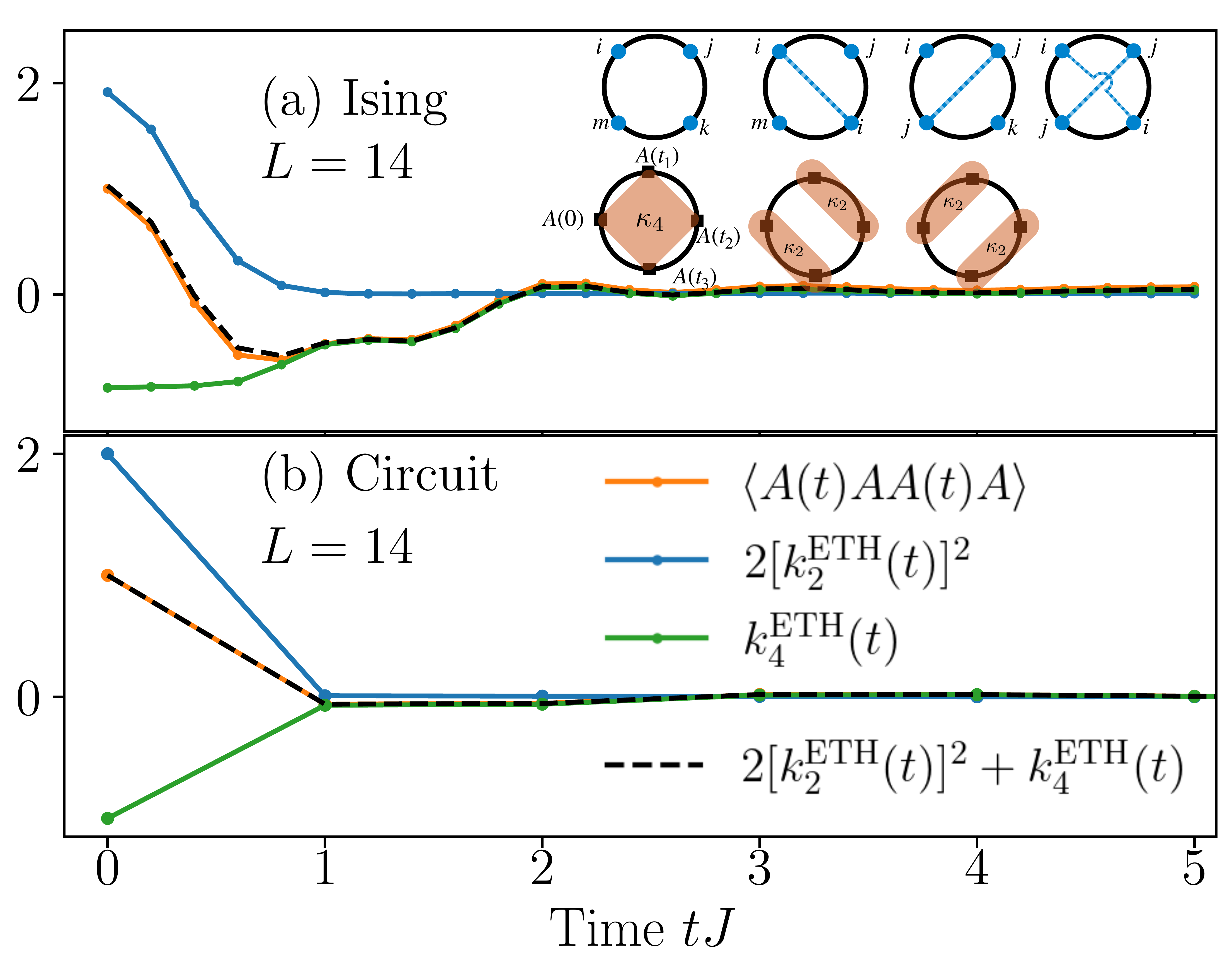}
	\caption{Multi-time correlation functions and ETH free-cumulants in chaotic systems. The full dynamics of $\langle \hat A(t) \hat A \hat A(t) \hat A\rangle$ (orange) is contrasted with the gaussian contribution $2[k^{\rm ETH}_2(t)]^2$ (blue), the free cumulant $k^{\rm ETH}_4(t)$ (green) and the sum of the two (dashed), cf. Eq.\eqref{eq_mom4}. (a) Ising model with $L=16$. (b) Floquet circuit with $L=14$.  Inset: decomposition in non-crossing and crossing partitions of $\langle \hat A(t_1) \hat A(t_2) \hat A(t_3) \hat A\rangle$ with $\langle \hat A\rangle=0$.  Top row: bookkeeping for the matrix elements $A_{ij}A_{jk}A_{km}A_{mi}$, the dashed line indicates an index contraction. Bottom row: non-crossing partitions in the dual partition and the associated free cumulants.   }
	\label{fig_1}
 \vspace{-.5cm}
\end{figure}
The high-order version of ETH was introduced in Ref.\cite{foini2019eigenstate} to describe correlation functions between $q$ times. 
 It predicts that the average of products of $q$ matrix elements with distinct indices is
\begin{equation}
    \label{ETHq}
    \overline{A_{i_1i_2}A_{i_2i_3}\dots A_{i_{q}i_1}} = e^{-(q-1)S(E^+)} F_{e^+}^{(q)}(\vec \omega) \ ,
\end{equation}
while products with repeated indices factorize in the large system size limit.
Here, $e^{-S(E^+)}$ is the mean level spacing at average energy $E^+=(E_{i_1}+\dots +E_{i_q})/q$,  
$\omega = (\omega_{i_1i_2}, \dots, \omega_{i_{q-1}i_q})$ with $\omega_{ij}=E_i-E_j$ are $q-1$ energy differences and $F_{e^+}^{(q)}(\vec \omega)$ is an order one smooth function of the energy density $e^+= E^+/N$ and $\vec \omega$, which encodes the physics. Throughout this paper, we will refer to Eq.\eqref{ETHq} as \emph{full ETH} \footnote{We remark that this is different from the generalization of ETH concerning integrable systems \cite{essler2016quench, leblond2020eigenstate}. }.
Building on this ansatz, 
Ref.\cite{pappalardi2022eigenstate} has recently argued that the building block of thermal multi-point correlation functions is given by \emph{free cumulants} $\kappa_q$, a type of connected correlation function defined in Free Probability (FP) \cite{speicher1997free}. 
This is a branch of math with applications in random matrices \cite{mingo2017free} and combinatorics \cite{nica2006lectures}, that has recently appeared in many-body physics \cite{pappalardi2022eigenstate, hruza2022dynamics}. 
However, to date, there has been no numerical investigation to test the full ETH in physically relevant many-body systems.

In this paper, we establish the validity of higher-order ETH \eqref{ETHq} by numerical investigation of free cumulants in generic many-body systems with finite local Hilbert space and local interactions. 
We consider two archetypical classes of chaotic systems believed to model generic many-body quantum dynamics, non-integrable spin chains and Floquet brickwork circuits.  

 Using exact numerical methods, we demonstrate the ETH properties for four-point correlations of local observables at infinite temperature as a function of time and frequency. We show that the dynamics of four-point time correlation functions {coincides at long times with} the fourth-order free cumulant, see Fig.~\ref{fig_1}. This holds due to the exponential suppression of crossing diagrams and the factorization of non-crossing ones, as predicted by ETH in the FP description. 
 To pinpoint the features of physical systems, we contrast ETH predictions with the results for rotationally invariant random matrices. While the latter and systems obeying full ETH share the same factorization for moments of matrices and multi-point averages, the important difference is in the shape of frequency-resolved cumulants, which is constant for random matrices but frequency-dependent in Hamiltonian and circuits.
To the best of our knowledge, this is the first numerical evaluation of the frequency-resolved higher-order ETH ansatz \eqref{ETHq}.
In the case of extensive operators (sum of local ones), the analysis shall be carried out in the microcanonical ensemble, as we discuss in the accompanying paper \cite{newpaper}.

{\bf{Free cumulants in ETH -}}  
{
The consequences of the full ETH \eqref{ETHq} are greatly simplified using the free cumulants defined in FP~\cite{voiculescu1985symmetries, speicher1997free}.
We focus on local observables and infinite temperature averages $\langle \bullet\rangle = \frac {\text{Tr}(\bullet)}{D}$ (where $D= \text{Tr}\, \mathbb 1$ is the Hilbert space dimension \footnote{Note that the Hilbert space dimension $D$ corresponds to the density of states at infinite temperature, i.e. ${D=e^{S(E_{\beta=0})}}$ in the thermodynamic limit.}) $\hat A$, for the general discussion see \cite{SM}.\\
Free cumulants $\kappa_q$ of order $q$ are a specific type of connected correlation function, which is defined recursively from moment-cumulant formula as
\begin{equation}
    \label{eq_freecumbeta}
    \langle \hat A(t_1) \hat A(t_2) \dots \hat A(t_q) \rangle = \sum_{\pi\in NC(q)} \kappa_\pi \left ( t_1, t_2, \dots  t_q\right )  \ ,
\end{equation}
where $\pi$ is a partition (decomposition of a set $\{1, \dots q\}$ in blocks) that sums over \emph{non-crossing partitions} $NC(q)$, where the blocks do not ``cross''.
In Eq.\eqref{eq_freecumbeta}, $\kappa_\pi(t_1, \dots, t_q) = \kappa_\pi \left ( A(t_1), \dots,  A(t_q) \right )$ are products of free cumulants, one for each block of $\pi$, see \cite{SM}.  At the first two orders, Eq.\eqref{eq_freecumbeta} yields $\braket{A}=\kappa_1$, $\braket{\hat A(t) \hat A(0)} = [\kappa_1]^2 + \kappa_2(t)$. For $q=4$ with $\kappa_1=\braket{\hat A}=0$, the $NC(4)$ are illustrated on the second line in the inset Fig.\ref{fig_1}a and $\kappa_4$ is defined implicitly by:
\begin{align}
\begin{split}
    \label{eq_mom4}
    \langle \hat A(t_1) & \hat A(t_2) \hat A(t_3) \hat A(0) \rangle 
     = \kappa^{}_4(t_1, t_2, t_3,0)  \\
     & \quad + \kappa^{}_2(t_1,t_2)\, \kappa^{}_2(t_3, 0)
    + \kappa^{}_2(t_2, t_3) \, \kappa_2^{}(t_1, 0)   \ .
\end{split}
\end{align}
Note that this is just an recursive definition of cumulants in terms of moments. In Ref.\cite{pappalardi2022eigenstate}, it was shown that 
the full ETH implies a compact and non-recursive expression for these free cumulants, given by summations with distinct indices of the matrix elements, i.e.
\begin{align}
    \label{freekETH}
    k^{\rm ETH}_{q}&
      ( t_1, \dots, t_q)   
    \\ & = 
    \frac 1D \sum_{i_1\neq i_2 \neq \dots \neq i_q}  A_{i_1i_2}A_{i_2i_3}\dots A_{i_{q}i_1}e^{i t_1 \omega_{i_1i_2} +\dots i t_q \omega_{i_{q}i_1}}\ ,\nonumber
\end{align}
This result is derived by showing that, due to ETH, the combinatorics of multi-time correlation functions and non-crossing partitions are the same. 
In fact, the correlators $\langle \hat A(t_1) \dots \hat A(t_q)\rangle$ are in principle obtained by summing over all indices the products of matrix elements $A_{i_1 i_2}\dots A_{i_qi_1}$. One can picture the latter via \emph{ETH diagrams} corresponding to matrix elements restricted over specific indices. Let us illustrate it for $q=4$: products $A_{ij}A_{jk}A_{km}A_{mi}$ are represented on a loop with four vertices/indices $i,j,k,m$, depicting energy eigenstates.  The contractions between indices are represented by lines that connect the vertices, see the first line in Fig.~\ref{fig_1}a. 
The ETH ansatz  \eqref{ETHq} implies that: 1) crossing partitions are exponentially (in $\log D$) subleading, e.g. 
\begin{align}
\label{eqUno}  
	       \frac 1D \sum_{i\neq j} 
        {|A_{ij}|^4} 
	       = \mathcal{O}(D^{-\alpha}) \ ,
        \quad 
\alpha > 0 \ ,        
\end{align}
and, 2) non-crossing partitions (also known as cactus diagrams \cite{foini2019eigenstate}) factorize, e.g.
\begin{align}
	\label{eqDue}
	 \frac{1}D &  \sum_{i\neq j \neq m} e^{\ii\omega_{ij} t_1 + i \omega_{im} t_2} {|A_{ij}|^2| A_{im}|^2} 
	 = k^{\rm ETH}_2(t_1) k^{\rm ETH}_2(t_2) 
	 \ ,
\end{align} 
for generic $t_1, t_2$ with $k^{\rm ETH}_2(t)$ given by Eq.\eqref{freekETH} for $q=2$. These two expressions imply that in the thermodynamic limit, the free cumulants defined implicitly in Eq.\eqref{eq_freecumbeta} correspond to the non-recursive result in Eq.\eqref{freekETH}, i.e. $\kappa_q(t_1, \dots t_q) \simeq k_q^{\rm ETH}(t_1, \dots t_q)$.}
\\
The ETH conditions \eqref{eqUno}-\eqref{eqDue} or equivalently Eq.~\eqref{freekETH} hold both in the time or the frequency domain. 
 By standard ETH manipulations \cite{SM}, for finite $\vec \omega= ( \omega_1,  \omega_2 ,  \omega_3)$, the Fourier transform of the free cumulants in Eq.~\eqref{freekETH} gives 
 \begin{equation}
 	\tilde k^{\rm ETH}_q(\vec \omega) = F^{(q)}_{e_{\beta=0}}(\vec \omega)\ ,
 \end{equation}
where $e_{\beta=0}$ is the infinite temperature energy. Thus, free cumulants yield all the physical properties as encoded in the smooth ETH functions of Eq.~\eqref{ETHq}.

We note that non-constant frequency dependence of free cumulants is the characteristic feature that distinguishes ETH from random matrix theory and not the absence of correlations. 
Generic \emph{rotationally invariant} $D\times D$ random matrices \cite{livan2018introduction} are characterized by probability distribution of the elements $P(A)\equiv P(A_{ij})$ that is invariant under a change of basis:
\begin{equation}
	P(A) = P(U^{-1} \, A \, U) \ ,
\end{equation}
where $U$ may be an orthogonal, unitary, or symplectic random matrix.  A class with this property is given by matrix models  $P(A) \propto \exp[- \frac D2 \text{Tr}\,V(A)]$. In the case of the Gaussian ensembles, $V(A)=A^2/2$, while a general anharmonic (e.g. polynomial) potential $V(A)$ yields a statistically correlated matrix $A$.
For rotationally invariant systems, Ref.\cite{maillard2019high} proved that the free cumulants are given by averages over simple loops (diagrams with different indices)
\begin{equation}
	\label{eq:FreeCumuRI}
	\,\overline{A_{i_1i_2}A_{i_2i_3} \dots A_{i_qi_1}  }  = D^{1-q}\,  \kappa_q(A) \ ,
\end{equation}
where $\kappa_q(A)$ is defined implicitly by cumulant-moment formula \cite{speicher1997free}, see also Ref.\cite{collins2007second}. 
Using the definition of the \emph{$q-$point spectral correlator} ${R_q(\vec \omega) = 	\sum_{i_1\neq i_2 \neq \dots \neq i_q} \overline{\delta(\omega_1 - \omega_{i_1 i_2})... }
 \overline{\delta(\omega_{q-1} - \omega_{i_{q-1} i_q})}}$, 
the frequency-dependent free cumulants read (see \cite{SM})
\begin{equation} 
	\label{eq:FreeCumuRIw}
	\tilde \kappa_q(\vec \omega) 
	 = \kappa_q(A) \frac{R_q(\vec \omega)}{D^{q}} \simeq \kappa_q(A)\ ,
\end{equation} where on the right-hand side we used $ \frac{R_q(\vec \omega)}{D^{q}}\simeq 1$ holding almost everywhere, i.e. when all frequencies $\omega_{ij}$ are much larger than mean level spacing \cite{mehta2004random}. Thus, the free cumulants in rotationally invariant random matrices are flat functions at finite frequencies.
Eq.~\eqref{eq:FreeCumuRIw} represents a factorization into a term depending only on the statistics of the eigenvalues ${R_q(\vec \omega)}/{D^{q}} $ and a constant depending only on the observable $A$. This has been referred to as spectral decoupling \cite{cotler2020spectral} and is a consequence of rotational invariance.


{\bf{Models -}}
As a paradigmatic example of chaotic non-integrable Hamiltonian, we consider the Ising chain with transverse and longitudinal fields described by 
\begin{equation}
\label{ising}
\hat H = \sum_{i=1}^L w \hat \sigma_i^x + \sum_{i=1}^L h \hat \sigma_i^z + \sum_{i=1}^{L}J \hat \sigma_i^z \hat \sigma_{i+1}^z \ ,
\end{equation}
where $\hat \sigma^\alpha_i$ is a Pauli matrix on site $i$ in the direction $\alpha = x,y,z$. 
We measure time in units of $J$ and set $w=\sqrt 5/2$, $h=(\sqrt 5+ 5)/8$. With periodic boundary conditions, this system has translation and space reflection symmetries. 
We focus on the local operator 
\begin{equation}
    \label{obsH}
	\hat A=\hat \sigma_{L/2}^z  \ ,
\end{equation}
in the middle of the spectrum, characterized by $\langle \hat A\rangle=0$. 
The results hold as well for other local observables at finite temperature with finite $\braket{A}_\beta\neq 0$, including the ones with support on two sites, as we show in \cite{SM}.

As a representative of chaotic Floquet evolution, we consider a brick-wall circuit design on a qubit chain of even length $L$ and periodic boundary conditions. {These models have attracted much attention recently, as they can be used to model discrete-time evolution in quantum computation and provide unique analytical insights \cite{BerKosPro2019b, bertini2018exact, claeys2020maximum, clayes2021ergodic}.} The system undergoes
discrete-time evolution governed by the unitary Floquet operator $\mathcal U$
with eigenvalues $e^{-\ii\phi_n}$,
defining quasi-energy spectrum $\{\phi_n\}\subset [0,2\pi)$
stepping in place of $\{E_n\}$.
Each time step is composed of two half-time steps such that 
$\mathcal U = \mathcal U_2 \mathcal U_1$ with 
\begin{equation}
    \mathcal U_1 =  U_1^{\otimes L/2} \ , \quad 
     \mathcal U_2 = \mathcal T \,  U_2^{\otimes L/2}\mathcal T^{-1},
\end{equation}
where $\mathcal T$ denotes a periodic shift on the lattice by one site. We show results for generic unitary two-qubit gates $U_1$ and $U_2$ (see \cite{SM}), with $U_1\neq U_2$ to avoid undesired symmetries. {The circuit lacks symmetry under spatial reflection, so we restrict ourselves to the full $k=0$ quasi-momentum subsector.} We fix a pair of randomly chosen gates but ensure that they are representative of generic gates. 
{We did all the calculations also for dual-unitary gates}, which were recently introduced as exactly solvable models of chaotic circuit dynamics \cite{BerKosPro2019b}, {for which we obtain the same results}. However, their defining space-time duality is not sufficient to allow for exact evaluation of the high-order ETH predictions and we hence {report the result on} generic gates only.
We consider the same single site-observable as in Eq.\eqref{obsH}.

 \begin{figure}[t]
	\centering
	\includegraphics[width=1 \linewidth]{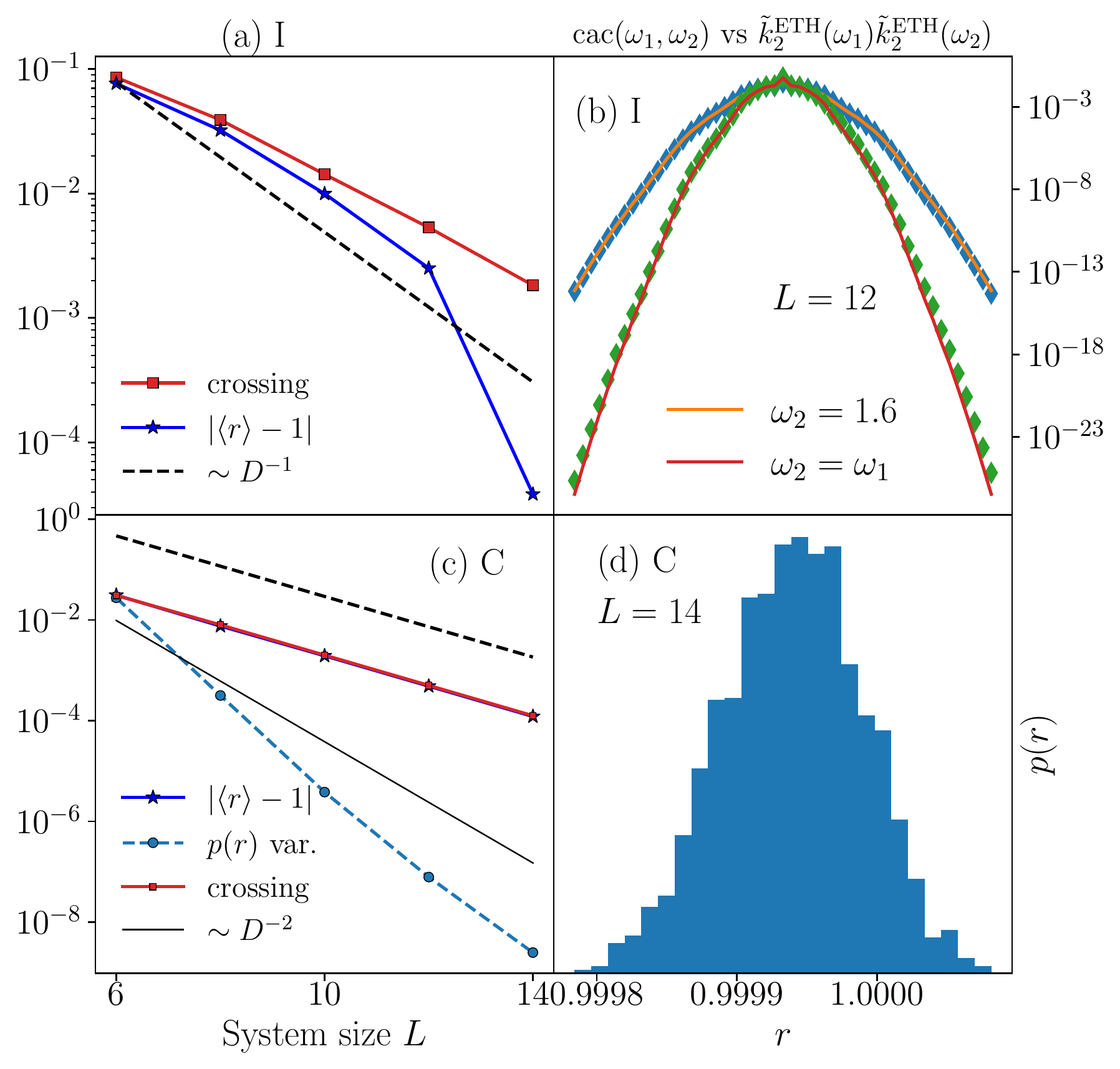}
	\caption{High-order ETH conditions \eqref{eqUno}-\eqref{eqDue} as a function of the system size $L$ in the Ising model (a) and the Unitary Circuit (c). {We denote with ``crossing'' Eq.\eqref{eqUno}, while $r$ corresponds to Eq.\eqref{eq_r}.} (b) Ising model factorization: the non-crossing contribution {$\text{cac}(\omega_1, \omega_2)$} (full line) in \eqref{eqDue} is compared with the factorized result {$\tilde k^{\rm ETH}_2(\omega_1)\tilde k^{\rm ETH}_2(\omega_2)$} (diamonds) in the frequency domain, along two directions $\omega_2 = 1.6$ and $\omega_2=\omega_1$ for $L=16$. (d) Distribution of the ratio $r$ with $\omega_1 \neq \omega_2$ in the circuit for $L=14$. The variance of this distribution as a function of $L$ is plotted in panel (c). 
 }
	\label{fig:2}
\end{figure}

{\bf{Results -}} 
We perform full diagonalization of $\hat H$ for the Ising model and $\mathcal U$ for the Circuit and test the full ETH ansatz by studying the statistics of $A_{nm}$ as a function of the energy $\omega_{nm}=E_n-E_m$ and the eigenphase differences $\omega_{nm}=\phi_n-\phi_m$, respectively. 
First, we look at the exponential suppression of crossing partitions [cf. Eq.~\eqref{eqUno}] and at the factorization of non-crossing parts [cf. Eq.~\eqref{eqDue}].
The contribution of the crossing partition  \eqref{eqUno} as a function of the system size $L$ is shown in Fig.\ref{fig:2}a,c for the Ising and Circuit models, respectively. Both cases show an exponential suppression with $L$, as the crossing contribution decreases with the inverse of the Hilbert space dimension $\sim D^{-1}$. 
Then, we check the factorization of the non-crossing partitions [cf. Eq.~\eqref{eqDue}]. To study it as a function of $L$, we consider the ratio
\begin{equation}
\label{eq_r}
   \langle  r \rangle = \frac {\text{cactus}(0, 0)}{[k^{\rm ETH}_2(0)]^2} \ ,
\end{equation}
where 
$\text{cactus}(t_1, t_2) = \frac{1}D \sum_{i\neq j \neq m} e^{\ii\omega_{ij} t_1 + i \omega_{im} t_2} {|A_{ij}|^2| A_{im}|^2}$
is the left-hand side of Eq.~\eqref{eqDue}. 
The full ETH predicts that the ratio in Eq.\eqref{eq_r} converges to one in the limit ${L\to \infty}$. This is shown in Fig.~\ref{fig:2}a,c, where we plot $|\langle r\rangle -1|$ for the Ising and the Circuit model, respectively. From the data of the Ising model, one can not infer the exact scaling with the system size. However, we note that the absolute values of $|\langle r\rangle-1|$ for the Circuit are at least one order of magnitude smaller than in the Hamiltonian case. This factorization holds also in the frequency domain, as shown in Fig.~\ref{fig:2}b,d. In panel b, we plot the Fourier transform of $\text{cactus}(t_1, t_2)$, i.e. 
\begin{eqnarray}
\hspace{-0.2 cm}
    \text{cac}(\omega_1, \omega_2)\!\!&=& \!\!\!\frac{1}D\!\!\!\sum_{i\neq j \neq m}\!\!\!
\delta_\tau(\omega_1\!-\!\omega_{ij})\delta_\tau(\omega_2\!-\!\omega_{im}) {|A_{ij}|^2| A_{im}|^2}
\nonumber
\end{eqnarray}
and the product $ \tilde k^{\rm ETH}_2(\omega_1)  \tilde k^{\rm ETH}_2(\omega_2)$ along two directions $\omega_2=\omega_1$ and $\omega_2=1.6$ for the Ising model with $L=16$. Here we indicate with $\delta_\tau(\omega)$ a smoothed delta function of width $1/\tau$, such that $\lim_{ \tau \to \infty} \delta_\tau(\omega) =\delta(\omega)$. We choose a Gaussian function $\delta_\tau(\omega)=\frac{\tau}{\sqrt{2\pi}} e^{- \tau^2 \omega^2 /2}$ of the continuous frequency, fixing $\tau\gg 1$ such that the results no longer depend on it ($\tau=2.5$ for the Ising model and $\tau=10$ for the circuit). The data show a good agreement between the two predictions. In panel d, we consider the Floquet circuit evolution and we show the distribution of the frequency-dependent ratio
$    r(\omega_1, \omega_2) = \frac{ {\text{cac}}(\omega_1, \omega_2)}{ \tilde k^{\rm ETH}_2(\omega_1)  \tilde k^{\rm ETH}_2(\omega_2)} $
for $L=14$ with $\omega_1\neq \omega_2$. 
The probability density $p(r)$ peaks at one, with a small variance that decreases with the system size with a double exponential $\sim D^{-2}$, as evidenced in Fig.~\ref{fig:2}c. The factorization of the non-crossing contributions holds also as a function of time, see \cite{SM}.
\begin{figure}[t]
	\centering
	\includegraphics[width=1 \linewidth]{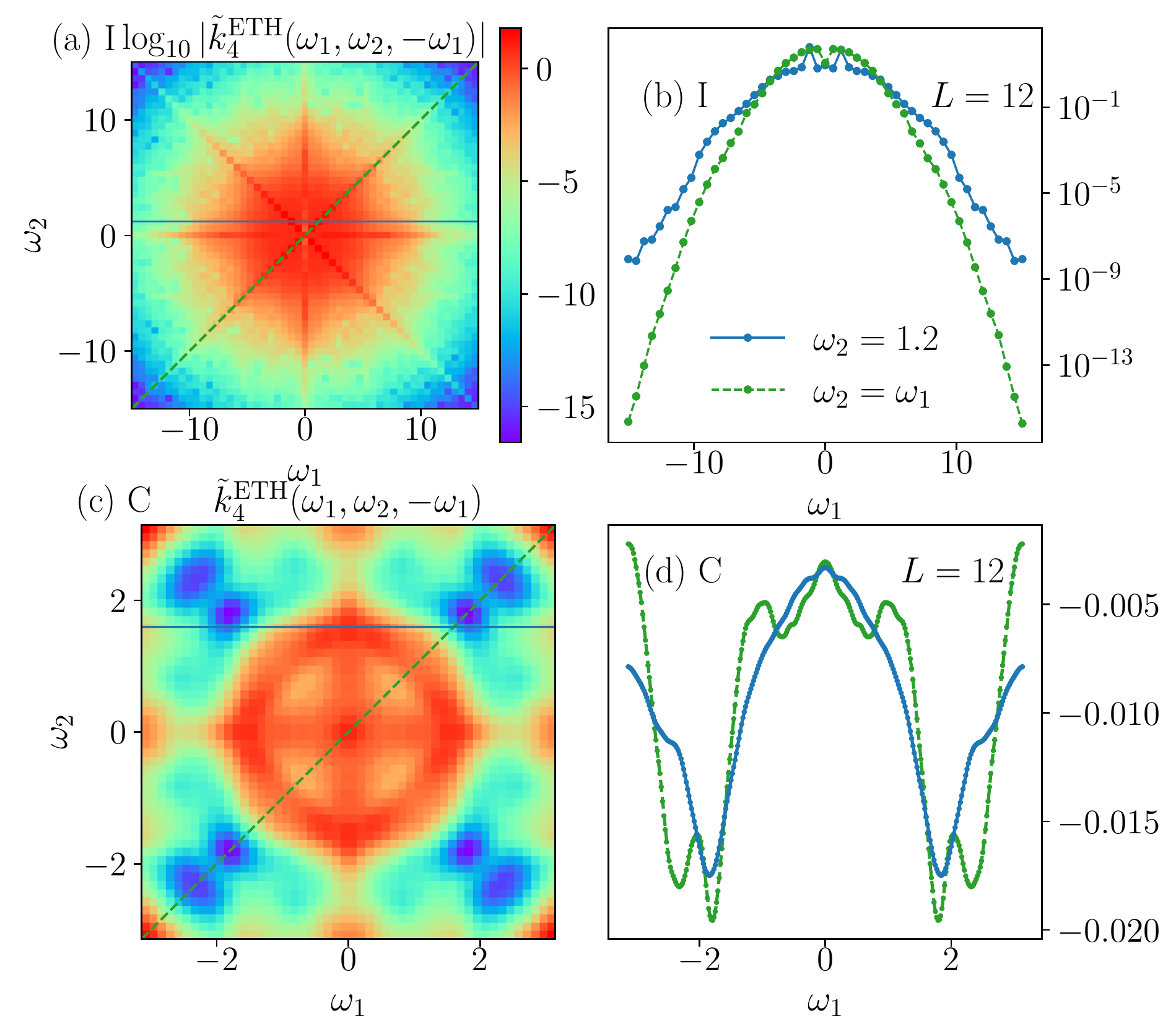}
	\caption{Frequency dependence of the fourth free cumulant $\tilde k^{\rm ETH}_4(\omega_1, \omega_2, -\omega_1)$ for the Ising model, with $L=16$ (ab) and the Floquet circuit with $L=12$ and $\tau=10$ (cd). On the right (b,d), the frequency dependence is plotted along two directions $\omega_1=\omega_2$ (green) and $\omega_2=1.6$ (blue). }
	\label{fig:3}
\end{figure}
The exponential suppression of crossing partitions and factorization of non-crossing ones lead to the decomposition in Eq.~\eqref{eq_mom4} with free cumulants given by the ETH prediction in Eq.\eqref{freekETH}. We test it by looking at the out-of-time order correlator (OTOC) $\langle \hat A(t) \hat A \hat  A(t) \hat  A\rangle $ for which it reads
\begin{equation}
    \label{ethOTOC}
    \langle \hat A(t) \hat A \hat A(t) \hat A\rangle \simeq 2 [k^{\rm ETH}_2(t)]^2 + k^{\rm ETH}_4(t)
\end{equation}
where $k^{\rm ETH}_4(t)$ is given by Eq.~\eqref{freekETH} with $t_1=t_3=t$ and $t_2=0$. The numerical results for the Ising and circuit models are presented in Fig.~\ref{fig_1}a,b for $L=16$ and $L=14$, respectively. The dynamics of $\langle \hat A(t) \hat A \hat A(t) \hat A\rangle $ is compared with the Gaussian (Wick) contribution $2[k^{\rm ETH}_2(t)]^2$, the fourth free cumulant $k^{\rm ETH}_4(t)$, and the sum of two [cf. Eq.~\eqref{ethOTOC}] which shall give the full result. The remarkable agreement for both classes of many-body evolution is one of the main results of the paper. 
After a short time, the two-point correlation $k^{\rm ETH}_2(t)$ approaches zero, while the OTOC displays slower dynamics, which is in turn well captured by $k^{\rm ETH}_4(t)$.
This shows that at long times all information about four-point functions are encoded in the fourth free cumulant $k^{\rm ETH}_4(t)$ which thus quantifies OTOC dynamics at longer time scales.

To characterize such higher-order correlations, we finally study the fourth free cumulant \eqref{freekETH} in frequency:
\begin{align}
    \label{free_cumuETH_W}
    \tilde k^{\rm ETH}_{4}(\omega_1, \omega_2, \omega_3) & = \frac 1D \sum_{i\neq j\neq m\neq k} 
        A_{ij}A_{jk}A_{km} A_{mi}
        \\ &  \times
    \delta_\tau(\omega_{ij} - \omega_1)\delta_\tau(\omega_{jk} - \omega_2)\delta_\tau(\omega_{km} - \omega_3) \ .\nonumber
\end{align} 
  The results are shown in Fig.~\ref{fig:3} for  $\omega_3=-\omega_1$. In panels (b) and (d) we plot the behaviour along two directions $\omega_2=\omega_1$ and $\omega_2=1.6$, for the Ising Hamiltonian and circuit respectively. Both classes display a clear frequency dependence of the fourth free cumulant $\tilde k^{\rm ETH}_4(\vec \omega)$. 
While in the Hamiltonian case $\tilde k^{\rm ETH}_4(\vec \omega)$ rapidly decays at large frequencies (Fig.~\ref{fig:3}ab), in the Floquet evolution frequencies have a period of $2 \pi$ and the non-constant frequency dependence is restricted to this interval (Fig.~\ref{fig:3}c,d).
This behaviour is thus in stark contrast with the result for random matrices (for which free cumulants are a flat function of frequency) [cf. Eq.~\eqref{eq:FreeCumuRIw}] and is the distinguishing feature of ETH for physical many-body lattice systems.

{\bf{Discussion and conclusions} -}
Our work illustrates that the general framework of ETH applies to local operators in non-integrable interacting lattice models.

Let us remark that dual-unitary Floquet circuits, representing a paradigm of maximally chaotic dynamics with local interactions~\cite{BerKosPro2019b}, 
display qualitatively exactly the same phenomenology. 
In particular, free cumulants depend on frequency, as already established for the
second cumulant $\tilde k^{\rm ETH}_2(\omega)$ in Ref.\cite{fritzsch2021eigenstate}. This behaviour differs from the structureless random matrix result of Ref.\cite{liao2022field} (cf. Eq.~\eqref{eq:FreeCumuRIw}) derived for locally interacting systems in the limit of infinite local Hilbert space dimension $d\to \infty$. 
Even though for finite $d$, dual-unitarity allows for analytically tractable results in the thermodynamic limit \cite{BerKosPro2019b}, establishing rigorous results in the context of ETH remains a challenging problem as it requires precise control of finite size corrections. For instance, even if time-order correlators can be understood via a transfer matrix approach, the dynamics of OTOC at finite times can be solved exactly only on the light cone \cite{claeys2020maximum}. Thus, it remains an open question whether one can obtain some analytical understanding of higher-order free cumulants for any type of finite $d$ models with local interactions. 

Our findings 
can be used as a starting point for further investigations. 
The behaviour of $k^{\rm ETH}_4(t)$ for the Ising model and the Floquet circuit is similar at long times (Fig.~\ref{fig_1}) and it is suggestive of some form of universality.
Such dynamical results indicate a hierarchy of correlation functions at different orders, $k^{\rm ETH}_2, k^{\rm ETH}_4$, etc, which should be of interest to explore systematically.

 Potential insights could be achieved by probing the full ETH conditions in the presence of integrability -- for which ETH is a highly debated open issue \cite{essler2016quench, Vidmar2019, Goold2020} -- but this is left for future research.

\medskip 

\begin{acknowledgments} 
We thank X. Turkeshi and G. Giudici for useful discussions, as well
as discussions and collaboration on related projects with L. Foini and J. Kurchan. We are grateful to A. Dymarsky for useful comments on the manuscript. 
S. P. has received funding from the European Union’s Horizon Europe program under the Marie Sklodowska Curie Action VERMOUTH (Grant No. 101059865). S.P. acknowledges support by the Deutsche Forschungsgemeinschaft (DFG, German Research Foundation) under Germany’s Excellence Strategy - Cluster of Excellence Matter and Light for Quantum Computing (ML4Q) EXC 2004/1 -390534769. F. F. acknowledges support by Deutsche Forschungsgemeinschaft (DFG), Project 
No. 453812159. T.P acknowledges support from Program P1-0402, and Grants N1-0219 and N1-0233 of the Slovenian Research and Innovation Agency (ARIS).

\end{acknowledgments}

\bibliography{biblio}

------------------------------
\widetext
\clearpage
\begin{center}
\textbf{\large \centering Supplemental Material:\\  }
\end{center}

\setcounter{equation}{0}
\setcounter{section}{0}
\setcounter{figure}{0}
\setcounter{table}{0}
\setcounter{page}{1}
\renewcommand{\theequation}{S\arabic{equation}}
\setcounter{figure}{0}
\renewcommand{\thefigure}{S\arabic{figure}}
\renewcommand{\thepage}{S\arabic{page}}
\renewcommand{\thesection}{S\arabic{section}}
\renewcommand{\thetable}{S\arabic{table}}
\makeatletter

\renewcommand{\thesection}{\arabic{section}}
\renewcommand{\thesubsection}{\thesection.\arabic{subsection}}
\renewcommand{\thesubsubsection}{\thesubsection.\arabic{subsubsection}}

\newcommand{\nocontentsline}[3]{}
\newcommand{\tocless}[2]{\bgroup\let\addcontentsline=\nocontentsline#1{#2}\egroup}

In this Supplementary Material, we provide additional analysis and background calculations to support the results in the main text. In Sec.\ref{app_freek}, we provide the theoretical background to understand free cumulants in ETH and for rotationally invariant random matrices.  In Sec.\ref{app_factot} we analyze the factorization of the cactus diagram in the time domain. In Sec.\ref{app_local}, we report the same results illustrated in the main text in the case of different local operators and in Sec.\ref{sec_finiteT} at finite temperature.
We conclude with the details on the circuit model in Sec.\ref{sec_circ}.

\tableofcontents

\section{Introduction to Free Cumulants and their expression in ETH}
\label{app_freek}
In this Appendix, we provide a self-contained and pedagogical introduction to the definition of free cumulants, starting from the combinatorial approaches to classical cumulants. Then, we  review their use to understand the full ETH Eigenstate Thermalization Hypothesis and finally discuss their behaviour in rotationally invariant systems.  \\

\subsection{Classical Cumulants}
\begin{figure}[t]
	\includegraphics[width=1\linewidth]{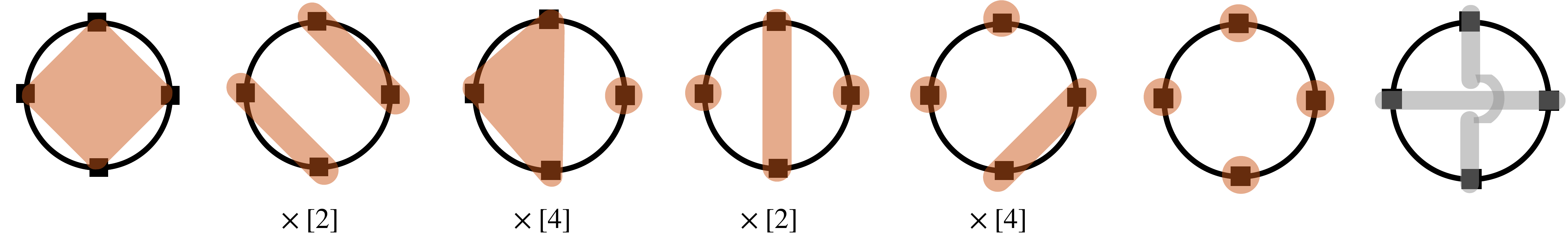}
    \label{fig_P4}
    \caption{Set of all partitions for $n=4$. With the colour orange, we represent the non-crossing partitions, while the crossing one is in grey. With $\times [m]$, we denote the $m$ cyclic permutations of that partition, which determines the coefficients appearing in the moment/cumulant formulas in Eq.\eqref{eq_clacu} and Eq.\eqref{eq_free_cumu_n}.}
 \end{figure}

Let us denote with ``classical cumulants'' the standard cumulants of commuting random variables. 
Consider $x$ a random variable with probability $p(x)$ and average $\mathbb E(\bullet)=\int \bullet p(x)dx $.
Classical cumulants $c_n(x)$ are defined as connected correlation functions: a suitable combination of moments $m_n = \mathbb E(x^n)$ of the same or lower order. For instance, the first four orders read
\begin{subequations}
\label{eq_clacu}
    \begin{align}
        c_1(x) & = \mathbb E(x)  \\
        c_2(x) & = \mathbb E(x^2) - \mathbb E(x)^2\\
        c_3(x) & = \mathbb E(x^3) - 3 \mathbb E(x^2)\mathbb E(x) + 2 \mathbb E(x)^3\\
        c_4(x) & = \mathbb E(x^4) - 3 \mathbb E(x^2)^2 - 4 \mathbb E(x)\mathbb E(x^3) + 12 \mathbb E(x^2)\mathbb E(x)^2 - 6 \mathbb E(x)^4 \ .\label{c4}
    \end{align}
\end{subequations}
Notably, the specific coefficients appearing in this expression can be obtained in a combinatorial way, based on the concept of partitions. A partition $\pi$ of a set $\{1, \dots n\}$ is a decomposition in blocks that do not overlap and whose union is the whole set. The set of all partitions of $\{1, 2, \dots n\}$ is denoted $P(n)$. The example of $P(4)$ for $\{1, 2, 3, 4\}$ is shown in Fig.\ref{fig_P4}, where with $\times [m]$ we denote that there are $m$ cyclic permutations. 
The number of the partitions of a set with $n$ elements is called the Bell number $B_n$ defined recursively as $B_{n+1} = \sum_{k=0}^n \binom{n}{k} B_k$ with $B_1=1$ and $B_2=2$, $B_3=5$, $B_4=15$, $B_5=52$, etc.
The classical cumulants \eqref{eq_clacu} can be defined implicitly by the moments/classical cumulants formula from the sum over all possible partitions
\begin{equation}
    \mathbb E(x^n) = \sum_{\pi \in P(n)} c_\pi(x) \quad \quad \text{with} \quad c_\pi(x) = \prod_{b\in \pi}c_{|b|}(x) \ ,
\end{equation}
where on the right-hand side $|b|$ denotes the number of elements of the block $b$ of the partition $\pi$. The result for the first four orders reads
\begin{subequations}
\label{eq_clamo}
    \begin{align}
    \mathbb E(x)  & =  c_1(x)  \\
    \mathbb E(x^2) & =  c_2(x) + c_1^2(x)\\
    \mathbb E(x^3) & = c_3(x) + 3 c_2(x) c_1(x) + c_1^3(x)    \\
     \mathbb E(x^4) & = c_4(x) + 3 c_2^2(x) + 6 c_2(x)c_1^2(x) + 4 c_3(x)c_1(x) + c_1^4(x) \ .
    \end{align}
\end{subequations}
Note that the coefficients correspond exactly to the multiplicities of each diagram. By inverting these relations one immediately finds the classical cumulants in Eq.\eqref{eq_clacu}. 

Here we only reported the definition for a single random variable, but the same can be easily extended to families of random variables $(x_1, x_2, \dots)$ from
\begin{equation}
    \mathbb E(x_1x_2\dots x_n) = \sum_{\pi \in P(n)} c_\pi(x_1x_2\dots x_n) \quad \text{with} \quad  c_\pi(x_1x_2\dots x_n) = \prod_{b\in \pi}c_{|b|}(x_{b(1)}x_{b(2)}\dots x_{b(n)})  \ ,
\end{equation}
where $b=(b(1), b(2), \dots , b(n))$ denotes the element of the block of the partition. \\

Summarizing, classical cumulants are connected correlation functions, whose coefficients can be computed from the combinatorial counting of partitions. A crucial property of Gaussian distributions is that cumulants of order greater than two vanish. 
Hence classical cumulants can be thought of as \emph{the} connected correlations such that $c_{n>2}=0$ for Gaussian random variables.

\subsection{Free cumulants}
We are now in the position to define free cumulants, which generalize the previous definition to \emph{non-commuting} variables. For definiteness, let us start by considering a $D\times D$ random matrix $A$ and the so-called ``expectation value'' 
\begin{equation}
    \phi(\bullet) = \lim _{D\to \infty}\frac 1D \mathbb E \left [\Tr (\bullet) \right ] \ ,
\end{equation}
which is well defined in the large $D$ limit and normalized, i.e. $\phi(\mathbb 1) = 1$. 

The definition of \emph{free} cumulants is based on the combinatorics of \emph{non-crossing partitions}, which are partitions that do not cross. The set of non-crossing partitions of $\{1, 2, \dots n\}$ is denoted by $NC(n)$ and enumerated by Catalan numbers $C_n = (1+1/n) \binom{2n}{n}$ with $C_1=1$, $C_2=2$, $C_3=5$, $C_4=14$, $C_5=42$, etc. Hence the number of crossing and non-crossing partitions differs from $n=4$ on, as shown in Fig.\ref{fig_P4}. Free cumulants $\kappa_n(A)$ are hence defined implicitly by
\begin{equation}
    \label{eq_qcum}
    \phi(A^n) = \sum_{\pi \in NC(n)} \kappa_\pi(A) \quad \text{with}\quad \kappa_\pi(A)= \prod_{b\in \pi} \kappa_{|b|}(A) \ ,
\end{equation}
where we recall that $|b|$ is the size of each block in the partition $\pi$. This expression for the first few orders reads
\begin{subequations}
    \label{eq_freemom}
    \begin{align}
        \phi(A) & = \kappa_1(A) \\
        \phi(A^2) & =  \kappa_2 (A)+\kappa_1(A)^2 \\
        \phi(A^3) & = \kappa_3(A) + 3 \kappa_2(A) \kappa_1(A) + \kappa_1(A)^3\\
        \phi(A^4) & =  \kappa_4(A)+ 2 \kappa_2(A)^2 + 6 \kappa_2(A) \kappa_1(A)^2 + 4 \kappa_3(A) \kappa_1(A) + \kappa_1(A)^4\ ,
    \end{align}
\end{subequations}
which, by inverting for $\kappa_n$ leads to
\begin{subequations}
	\begin{align}
		\kappa_1(A) & = \phi(A)\ ,\\
		\kappa_2(A) & = \phi(A^2) - \phi(A)^2\ ,		\label{k2}\\
		\kappa_3(A) & = \phi(A^3)  - 3 \phi(A^2) \phi(A)+2 \phi(A)^3\ ,		\\
		\kappa_4(A) & = \phi(A^4) - 2\phi(A^2)^2 - 4 \phi(A)\phi(A^3) + 10\phi(A^2)\phi(A)^2  - 5 \phi(A)^4 \ .\label{k4_momr}
	\end{align}
\end{subequations} 
The first difference between classical and free cumulants appears in the fourth order, as one notices by comparing the factor $2 \times \phi(A^2)^2 $ in Eq.\eqref{eq_freemom} instead of the $3 \times \mathbb E(x^2)^2$ in Eq.\eqref{eq_clacu}.

For Gaussian random \emph{matrices}, the free cumulants of order greater than two vanish. It is now clear that free cumulants are the direct generalization of classical cumulants to non-commuting objects and that they can be thought of as \emph{the} connected correlations such that $\kappa_{n>2}=0$ for Gaussian random matrices.\\

The definition of free cumulants can be immediately extended to different random matrices $(A^{(1)}, A^{(2)}, ...)$ as 
\begin{equation}
    \label{eq_free_cumu_n}
    \phi(A^{(1)}A^{(2)}...A^{(n)}) = \sum_{\pi \in NC(n)} \kappa_\pi(A^{(1)}A^{(2)}...A^{(n)}) \quad \text{with} \quad  \kappa_\pi(A^{(1)}A^{(2)}...A^{(n)}) = \prod_{b\in \pi}k_{|b|}(A^{(b(1))}A^{(b(2))}...A^{(b(n))})  \ ,
\end{equation}
where $b=(b(1), b(2), \dots , b(n))$ denotes the element of the block of the partition. As an example, consider the following partition $\pi$ for $n=8$:

\begin{figure}[H]
\centering
\includegraphics[width=.2\linewidth]{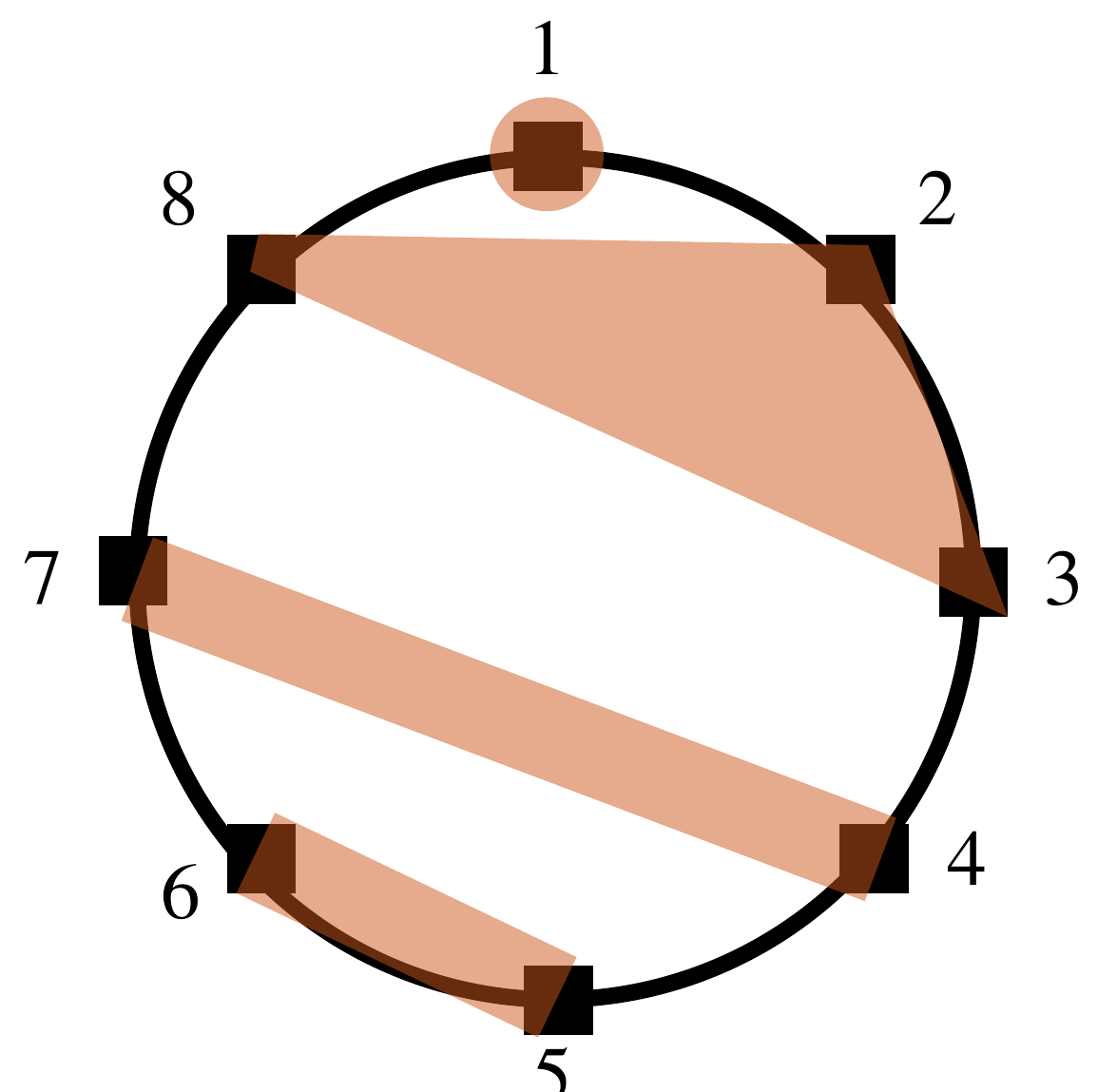}
\caption{A non-crossing partition $\pi$ of $n=8$.}
\label{fig_prodB}
\end{figure}

Here the the partition is $\pi = \{\{1\}, \{2,3,8\}, \{4,7\}, \{5, 6\}\}$ and the corresponding contribution reads:
\begin{equation}
    \kappa_\pi(A^{(1)}... A^{(8)}) = \kappa_1(A^{(1)}) \kappa_3(A^{(2)}A^{(3)}A^{(8)})\kappa_2(A^{(4)}A^{(7)})\kappa_2(A^{(5)}A^{(6)}) 
    \ .
\end{equation}

By inverting the implicit definition in Eq.\eqref{eq_free_cumu_n}, the first few free cumulants read
\begin{subequations}
	\begin{align}
		\kappa_1(A^{(1)}) & = \phi(A^{(1)})\ ,\\
		\kappa_2(A^{(1)}A^{(2)}) & = \phi(A^{(1)}A^{(2)}) - \phi(A^{(1)})\phi(A^{(2)})\ ,		\label{k2_}\\
  \begin{split}
      	\kappa_3 (A^{(1)}A^{(2)} A^{(3)})& = \phi(A^{(1)}A^{(2)} A^{(3)})  
       -  \phi(A^{(1)}A^{(2)}) \phi(A^{(3)}) 
       - \phi(A^{(1)}A^{(3)})\phi( A^{(3)})  
       - \phi(A^{(2)} A^{(3)})\phi(A^{(1)})
  \\ & \quad + 2 \phi(A^{(1)}) \phi(A^{(2)}) \phi(A^{(3)})\ .
  \end{split}
	\end{align}
\end{subequations} 

\subsection{Full ETH and Free cumulants}

\begin{figure}[b]
	\includegraphics[width=1\linewidth]{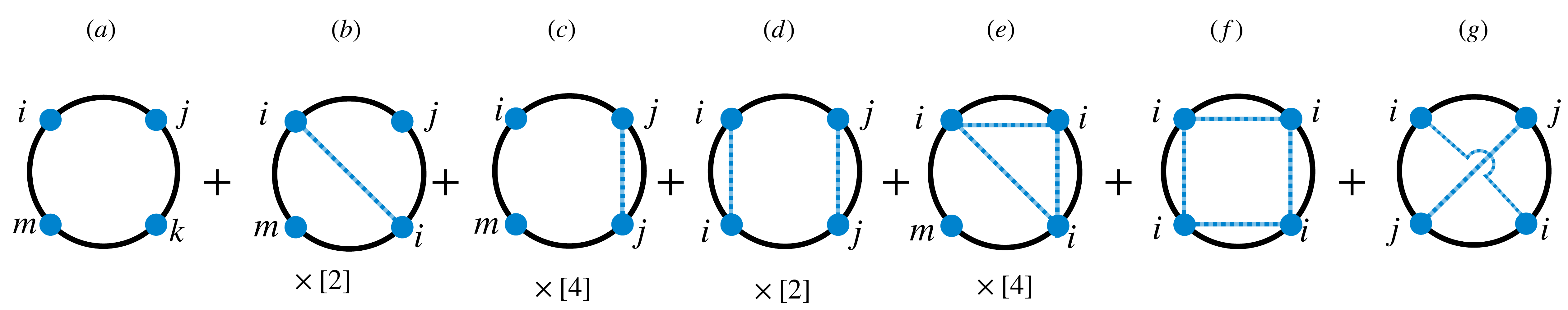}
    \label{fig_ethDiag}
    \caption{Bookeeping of ETH matrix elements for $n=4$. Matrix elements $A_{ij}$ lie on the vertex connecting two dots, which represent the energy index. Blue dots represent different indices and the edges connecting two or more dots represents a contraction among them. }
 \end{figure}
In this Appendix, we review the Free Probability approach to the full ETH ansatz as discussed in Ref.\cite{pappalardi2022eigenstate}.\\

First of all, revisiting the definition \eqref{eq_free_cumu_n}, one can \emph{define} recursively thermal free cumulants $k^\beta_q$ by
\begin{equation}
    \label{eq_freecumbeta_app}
    \langle A(t_1) A(t_2) \dots A(t_q) \rangle_\beta = \sum_{\pi\in NC(q)} \kappa^\beta_\pi \left ( A(t_1) A(t_2) \dots  A(t_q) \right )  \ ,
\end{equation}
where 
$\langle \bullet \rangle_\beta = \frac 1Z \Tr \left ( e^{-\beta H} \bullet \right ) $
with $ Z=\Tr \left ( e^{-\beta H} \right )$ plays the role of the expectation value $\phi()$.
Here $\kappa^\beta_\pi$ are products of thermal free cumulants one for each block of $\pi$, i.e.
\begin{equation}
     \kappa^\beta_\pi \left ( A(t_1) A(t_2) \dots  A(t_q) \right ) = \prod_{b\in \pi} \kappa^\beta_{|b|} (A(t_{b(1)}) A(t_{b(2)})\dots A(t_{b(n)})) \ ,
\end{equation}
where $|b|$ is the size of the block $b$ in the partition $\pi$. See the example in Fig.\ref{fig_prodB}, where now $t_{b(j)}$ keeps track of $b(j)$ in keeping track of the element of the block of the partition.
{
In what follows we will use the notations
$\kappa^\beta_\pi \left ( t_1, t_2, \dots t_q \right ) = \kappa^\beta_\pi \left ( A(t_1) A(t_2) \dots  A(t_q) \right )$
and \emph{time translational symmetry}, i.e. $\langle A(t_1) A(t_2) \dots A(t_q) \rangle_\beta=\langle A(t_1-t_q) A(t_2-t_q) \dots A(0) \rangle_\beta$ and  $\kappa^\beta_\pi \left ( t_1, t_2, \dots t_q \right ) = \kappa^\beta_\pi \left ( t_1-t_q, t_2-t_q, \dots 0 \right )$. For $q=1$ this reads $\kappa_1^{\beta}(t)=\kappa_1^\beta(0)= \kappa_1^\beta \equiv \braket{A}_\beta$.
}
Exactly as in Eq.\eqref{eq_free_cumu_n}, this is just an implicit definition of cumulants in terms of moments, which can be defined in principle also for integrable or non-ergodic systems. 
We will now discuss how this definition simplifies the discussion of the full ETH which, in turn, implies a particularly simple form for the thermal free cumulants.

The full version of the Eigenstate-Thermalization-Hypothesis has been introduced by Ref.\cite{foini2019eigenstate} as an ansatz on the statistical properties of the product of $q$ matrix elements $A_{ij}$. Specifically, the average of products with distinct indices $i_1\neq i_2 \dots \neq i_q$ reads
\begin{subequations}
    \label{GEN_ETH_app}
\begin{equation}
    \label{ETHq_app}
    \overline{A_{i_1i_2}A_{i_2i_3}\dots A_{i_{q}i_1}} = e^{-(q-1)S(E^+)} F_{e^+}^{(q)}(\omega_{i_1i_2}, \dots, \omega_{i_{q-1}i_q}) 
\end{equation}
while, with repeated indices, it shall factorize in the large $N$ limit as
\begin{align}
\label{ETH_conta_app}
 \overline{A_{i_1i_2}\dots A_{i_{k-1}i_1}A_{i_1i_{k+1}}\dots A_{i_{q}i_1}}
 =  \overline{A_{i_1i_2}\dots A_{i_{k-1}i_1}} \; 
\overline{A_{i_1i_{k+1}}\dots A_{i_{q}i_1}}    \ .    
\end{align}
\end{subequations}
In Eq.\eqref{ETHq}, $Ne^+ = E^+=(E_{i_1}+\dots +E_{i_q})/q$ is the average energy, $\vec 
\omega = (\omega_{i_1i_2}, \dots, \omega_{i_{q-1}i_q})$ with $\omega_{ij}=E_i-E_j$ are $q-1$ energy differences and $F_{e^+}^{(q)}(\vec \omega)$ is a smooth function of the energy density $e^+=E^+/N$ and $\vec \omega$. Thanks to the explicit entropic factor, the functions $F^{(q)}_E(\vec \omega)$ are of order one and they contain all the physical information.\\

One would like to understand how this ansatz applies to multi-time correlation functions as
\begin{equation}
\label{eq_Aq}
    \langle A(t_1) A(t_2) \dots A(0) \rangle_\beta \ .
    \end{equation}
The full result is given by the sum over all indices of ${A_{i_1i_2}A_{i_2i_3}\dots A_{i_{q}i_1}}$. Thus, to determine the contribution of the different matrix elements one shall consider all the possible contractions. 
One can do this \emph{diagrammatically}. Let us consider the example of four-point functions that we illustrate pictorially in Fig.\ref{fig_ethDiag}.  Products $A_{ij}A_{jk}A_{km}A_{mi}$ are represented on a loop with four vertices $i,j,k,m$, depicting energy eigenstates.  The contractions between two or more indices are represented by lines that connect the vertices. The blue dots indicate that the indices are all different. For instance, the first diagram represents $A_{ij}A_{jk}A_{km}A_{mi}$, the second $A_{ij}A_{ji}A_{im}A_{mi}$, the third $A_{ij}A_{jj}A_{jm}A_{mi}$  with all distinct indices, and so on.  One recognizes that there are two types of diagrams: (1) \emph{non-crossing} ones -- in which the polygons created by the indices do not cross (the simple loops with all different indices are in this class) -- and (2) \emph{crossing ones}, in which the lines cross. The full ETH ansatz in Eqs.\eqref{GEN_ETH} in the large $L$ limit implies that
\begin{enumerate}
    \item all non-crossing diagrams yield a finite contribution  with \emph{factorization of non-crossing diagrams into products of irreducible simple loops}:
    \begin{equation}
        \label{eq_noncrossing}
        \frac{1}Z  \sum_{i\neq j \neq m} e^{-\beta E_{ii}}e^{\ii\omega_{ij} t + i \omega_{im} t'} {|A_{ij}|^2| A_{im}|^2} 
	 \simeq 
	 \left ( \frac{1}Z  \sum_{i\neq j} e^{-\beta E_{ii}}e^{-i\omega_{ij} t }|A_{ij}|^2
  \right )
  \left (\frac{1}Z  \sum_{i\neq m} e^{-\beta E_{ii}}e^{-i\omega_{im} t'} |A_{im}|^2
  \right ) + O(N^{-1})\ ,
    \end{equation}
    for every $t, t'$. This property means that the diagrams (b-f) in Fig.\ref{fig_ethDiag} can be cut along the blue line. Such factorization holds in the large $L$ limit and is derived solving the left-hand side via saddle-point \cite{pappalardi2022eigenstate}.
    \item Crossing diagrams are suppressed with the inverse of the density of states as
      \begin{equation}
      \label{full_ETH2}
        \frac{1}Z  \sum_{i\neq j} e^{-\beta E_{ii}}e^{\ii\omega_{ij} t} {|A_{ij}|^4} 
	 \simeq e^{- S(E)} \sim D^{-1}\ ,
    \end{equation}
    for every time $t$. This property means that they can be neglected for computing higher-order correlation functions. 
\end{enumerate}

Eqs.\eqref{eq_noncrossing}-\eqref{full_ETH2} show that the combinatorics of multi-time correlation functions is the same as the one known in Free Probability:
all the contribution to multi-time correlations has to be found in non-crossing partitions. Specifically, the non-crossing ETH diagrams ((a-f) in Fig.\ref{fig_ethDiag}) can be read as the ``dual''  of non-crossing partitions $\pi$ in which every element of the set is not associated with an observable [(a-f) in Fig.\ref{fig_P4}]. Furthermore, the ETH factorization in Eq.\eqref{full_ETH2} implies a particularly simple form for the free cumulant $\kappa^{\beta}_{q}$ defined recursively in Eq.\eqref{eq_freecumbeta}, 
and it leads to the main result of the Free Probability approach to ETH:
The thermal free cumulants of ETH-obeying systems are given only by summations with distinct indices \cite{pappalardi2022eigenstate}
\begin{align}
    \label{freekETH}
    k^{\rm ETH}_{q} \left (  t_1, t_2, \dots, t_q \right )  & = 
    \frac 1Z \sum_{i_1\neq i_2 \neq \dots \neq i_q} e^{-\beta E_{i_1}} A_{i_1i_2}A_{i_2i_3}\dots A_{i_{q}i_1}e^{i t_1 \omega_{i_1i_2} +\dots i t_q \omega_{i_{q}i_1}}\\
    & = \text {FT} \left [ F^{(q)}_{e_\beta}(\vec \omega) e^{-\beta \vec \omega \cdot \vec \ell_q} \delta(\omega_1 +\omega_2 + \dots +\omega_q)\right ] \ , \label{freekETH_b}
\end{align}
where in the second line $\text {FT} [\bullet]= \int d\vec \omega e^{-i \vec \omega \cdot \vec t} \bullet$ is the Fourier transform and $e_\beta = \langle H\rangle_\beta/N$ is the thermal energy density. {For $q=1$ this reads $k_1^{\rm ETH} (t) = k_1^{\rm ETH} (0)=k_1^{\rm ETH} = \braket{A}_\beta$.} {To keep the discussion self-consistent, we report the derivation of Eq.\eqref{freekETH_b} below. } The thermal weight with $\vec \ell_q = \left (\frac{q-1}{q}, \dots, \frac 1q , 0\right )$ corresponds to a generalization of the fluctuation-dissipation theorem.
\emph{This result shows that all the correlations of the full ETH \eqref{ETHq} are encoded precisely in the thermal free cumulants. }
On four-point functions, the validity of the ETH ansatz implies:
\begin{align}
    \label{eq_20}
    \begin{split}
        \langle A(t_1) A(t_2) A(t_3) A(t_4) \rangle_\beta = & 
        k_4^{\rm ETH}(t_1, t_2, t_3, t_4)  
        + k_2^{\rm ETH}(t_1,t_2) k^{\rm ETH}_2(t_3, t_4)
        + k_2^{\rm ETH}(t_1,t_4) k^{\rm ETH}_2(t_2, t_3)
        \\ & 
        + k^{\rm ETH}_1 \Big [
        k^{\rm ETH}_3(t_1, t_2, t_3)  
        + k^{\rm ETH}_3(t_1, t_3, t_4) 
        + k^{\rm ETH}_3(t_1, t_2, t_4) 
        + k^{\rm ETH}_3(t_2, t_3, t_4) \\
        & \quad\quad + k^{\rm ETH}_1 k^{\rm ETH}_2(t_1, t_3)
        + k^{\rm ETH}_1 k^{\rm ETH}_2(t_2, t_4) + (k^{\rm ETH}_1)^3 
        \\
        & \quad \quad 
       +  k^{\rm ETH}_1 k^{\rm ETH}_2(t_1, t_2)  
        + k^{\rm ETH}_1 k^{\rm ETH}_2(t_1, t_4) 
        + k^{\rm ETH}_1 k^{\rm ETH}_2(t_2, t_3) 
        + k^{\rm ETH}_1 k^{\rm ETH}_2(t_3, t_4)
        \Big ]\ .
    \end{split}
\end{align}
For $k_1=0$, this expression reduces to Eq.(15) of the main text, where we 
used for short-hand notations $k^{\rm ETH}_q(t_1-t_q, t_2-t_q, ..., t_{q-1}-t_q) = k^{\rm ETH}_q(t_1-t_q, t_2-t_q, ..., t_{q-1}-t_q, 0)$.\\

{\bf Derivation of Equation \eqref{freekETH_b} - }
For the sake of completeness, let us recall here some ``standard ETH calculations'', see e.g. Ref.\cite{srednicki1999approach} or Ref.\cite{dalessio2016from}. In particular, we report the derivation of Eq.\eqref{freekETH_b} from Eq.\eqref{freekETH}, see Ref.\cite{pappalardi2022eigenstate}. Let us use translational invariance and consider $k^{\rm ETH}_q(t_1, t_2, ... , t_{q-1}, t_q=0)$ in Eq.\eqref{freekETH} as given by
\begin{align}
    k_q^{\rm ETH}(t_1, t_2, \dots t_{q-1}, 0) & = \frac 1Z \sum_{i_1 \neq i_2 \dots \neq i_q }e^{-\beta E_{i_1}} e^{i t_1(E_{i_1}-E_{i_2}) +  t_2(E_{i_2}-E_{i_3}) + \dots  t_{q-1}(E_{i_{q}}-E_{i_{q-1}})} 
    {A_{i_1i_2}A_{i_2 i_3} \dots A_{i_{q}i_1}} 
    \\ & =
    \frac 1Z \sum_{i_1 \neq i_2 \dots \neq i_q }e^{-\beta E_{i_1}} e^{i \vec t \cdot \vec \omega} e^{-(q-1) S(E^+)}  F^{(q)}_{e^+}(\vec \omega) \\
    & =
    \frac 1Z \int dE_1 \dots dE_q e^{-\beta E_1 } e^{i \vec t \cdot \vec \omega} e^{S(E_1) + \dots S(E_q) - (q-1) S(E^+)}\, F^{(q)}_{e^+}(\vec \omega)
\end{align}
where from the first to the second line we have used $\vec 
\omega = (\omega_{i_1i_2}, \dots, \omega_{i_{q-1}i_q})$ with $\omega_{ij}=E_i-E_j$ and substituted the ETH ansatz [c.f. Eq.\eqref{ETHq}] and from the second to the third we have exchanged the summation with the integral $\sum _{i_1}\to \int dE_1 e^{S(E_1)}$. 
We can thus Taylor expand the entropies around energy $E^+$ as
\begin{equation}
    S(E_i) = S(E^+ + (E_i-E^+)) = S(E^+) + S'(E^+)(E_i - E^+) + \frac 12 S''(E^+)(E_i - E^+)^2 + \dots \ .
\end{equation}
Then, by summing over all the energies one obtains
\begin{equation}
    \label{eq_expEntro}
    \sum_{i=1}^q S(E_i) = q S(E^+) +  S''(E^+) \sum_i (E_i - E^+)^2 + \dots \ ,
\end{equation} 
where the linear term in $E_i - E^+$ vanishes (due to $E^+ = (E_1 + E_2 + \dots E_q)/q$), while the quadratic term is subleading due to the thermodynamic property $S''(E^+) = -\beta^2 /C$ with $C\propto N$ the heat capacity and $\beta = S'(E^+)$ the inverse temperature at energy $E^+$. Since $E_i - E^+ \propto \vec \omega$ and $F_{e^+}(\vec \omega)$ is a smooth function that decays decays fast at large frequencies, we can neglect the second term in Eq.\eqref{eq_expEntro}. The free cumulant then reads
\begin{align}
    \label{eq:22}
      k_q^{\rm ETH}(\vec t) = \frac 1Z \int dE_1 e^{-\beta E_1 + S(E^+)} \int dE_2 \dots dE_{E_q} e^{i \vec t \cdot \vec \omega} F^{(q)}_{e^+}(\vec \omega)   \ .
\end{align}
We can now rewrite 
\begin{align}
      E_1 & = E^+ + (E_1 - \frac{E_1 + E_2 + \dots E_q}q) = E^+ + \frac{q-1}{q}  (E_1 - E_2) + \frac{q-2}{q} (E_2-E_3) + \dots + \frac 1q (E_{q-1}-E_q) 
      \\ & = E^+ + \vec \ell_{q} \cdot \vec \omega \ ,
\end{align}
where we have defined the ladder vector 
\begin{equation}
    \label{eq:ladderQ}
    \vec \ell_q = \left ( \frac{q-1}{q}, \frac{q-2}{q} \dots , \frac{1}{q}\right ) \ .
\end{equation}
We substitute this into Eq.\eqref{eq:22} and change integration variables $dE_1 dE_2 \dots dE_1 = dE^+ d\omega_1 d\omega_1 \dots d\omega_{q-1} $, leading to
\begin{align}
    k_q^{\rm ETH}(\vec t) = \frac 1Z \int dE^+ e^{-\beta E^+ + S(E^+)}
    \int d \omega_1 \dots d\omega_{q-1} e^{i \vec t \cdot \vec \omega - \beta \vec \ell_q \cdot \vec \omega } F^{(q)}_{e^+}(\vec \omega) \ .
\end{align}
Since $F^{(q)}_{e^+}(\vec \omega)$ is a smooth function of $E^+$ of order one, we can solve the integral over $E^+$ by saddle-point, which simplifies with the denominator and fixed the energy by the thermodynamic definition via $S'(E_\beta) = \beta$. This immediately leads to the desired Eq.\eqref{freekETH_b}.

\subsection{Free cumulants in rotationally invariant systems}
\label{appRI}
In this Appendix, we derive the result in Eq.(10) of the main text.\\

 \emph{Rotationally invariant} models are characterized by probability distribution of the matrix elements $P(A)\equiv P(A_{ij})$ that is invariant under a change of basis 
\begin{equation}
    \label{pa}
	P(A) = P(U^{-1} \, A \, U) \ ,
\end{equation}
where $U$ may be an orthogonal, unitary, or symplectic matrix {(see Section 3.2 of Ref.\cite{livan2018introduction})}. Let us denote $\overline{\bullet}$ averages over these ensembles. A class that enjoys this property is given by $P(A) \propto \exp[- \frac D2 \text{Tr}V(A)]$ where $V(A)$ is a generic polynomial - the potential ($V(A)=A^2/2$ in the case of the Gaussian ensemble). These matrices have the property that their moments only depend on the distributions of the eigenvalues $a_i$, i.e. $\langle A^m\rangle = \frac 1D \Tr(A^m) = \frac 1D \sum_i a_i^m$.

For rotationally invariant systems, Ref.~\cite{maillard2019high} proved that the free cumulants are given by averages over simple loops (diagrams with different indices)
\begin{equation}
	\label{eq:FreeCumuRI_sup}
	\,\overline{A_{i_1i_2}A_{i_2i_3} \dots A_{i_qi_1}  }  = D^{1-q}\,  \kappa_q(A) \quad \text{with} \quad i_1 \neq \dots \neq i_q
\end{equation}
at the leading order in $D$. Here the average is taken according to $P(A)$ in Eq.~\eqref{pa} and $\kappa_q(A)$ is defined in Eq.~\eqref{eq_qcum}, {with respect to the expectation value $\phi(\cdot) = \lim_{D\to \infty} \frac {\overline{\rm Tr(\cdot)}}D$.} The equality holds for each product of matrix elements, without the summation [in contrast with Eq.~(5) of the main text].
The overall constant normalization $D^{q-1}$ stems from the fact that the average over each element is the same.
For $q=2$ this equation reads
\begin{equation}
	\overline{A_{ij}A_{ji}} = D^{-1}\, \kappa_2(A) = D^{-1}\,\left (\, \langle A^2\rangle - \langle A\rangle^2\, \right ) \ ,
\end{equation}
where $\kappa_2(A)$ is given in Eq.~\eqref{k2}. This result corresponds to the ETH ansatz for $q=2$ for GUE (CUE) matrices.

We now compute free cumulants in the frequency domain defined as
\begin{equation}
	\label{eq:freeOmega}
	\tilde \kappa_q(\omega_1, \omega_2, \dots, \omega_{q-1}) = \frac 1D
	\sum_{i_1\neq i_2 \neq \dots \neq i_q}
	\overline{A_{i_1i_2}A_{i_2i_3} \dots A_{i_qi_1} 
	\delta(\omega_1 - (E_{i_1}-E_{i_2}))
	\dots 
	\delta(\omega_{q-1} - (E_{i_{q-1}}-E_{i_q}))} \ .
\end{equation}
We \emph{assume} a decoupling between the average over ${A_{i_1i_2}A_{i_2i_3} \dots A_{i_qi_1}}$ and the average over the delta functions 
(small statistical correlation between matrix elements and the spectrum). Substituting Eq.~\eqref{eq:FreeCumuRI_sup} into Eq.~\eqref{eq:freeOmega} we thus conclude
\begin{equation}
	\label{boxed}
	\tilde \kappa_q(\omega_1, \omega_2, \dots, \omega_{q-1}) 
	 = \kappa_q(A) \frac{R_q(\omega_1, \omega_2, \dots \omega_{q-1})}{D^{q}} \ ,
\end{equation}
where 
\begin{equation}
R_q(\omega_1, \omega_2, \dots \omega_{q-1}) = 	\sum_{i_1\neq i_2 \neq \dots \neq i_q} \overline{
\delta(\omega_1 - (E_{i_1}-E_{i_2}))
\delta(\omega_2 - (E_{i_2}-E_{i_3}))
\dots
\delta(\omega_{q-1} - (E_{i_{q-1}}-E_{i_q}))}
\end{equation}
is the \emph{$q-$point spectral correlator}, which encodes all $q-$point correlations between the energy eigenvalues. This generalizes to many $q$ the connected two-point spectral correlation $R_2(\omega) = \sum_{i\neq j} 
\delta(\omega -\omega_{ij})$, whose Fourier transform yields the spectral form factor. \\

For $q=2$, Eq.\eqref{eq:FreeCumuRI_sup} gives $\overline{A_{ij}A_{ji}}  = D^{-1}\, \kappa_2(A) $ and this result reads
\begin{equation}
\tilde \kappa_2(\omega) = \kappa_2(A) \frac{R_2(\omega)}{D^2} \ ,
\end{equation}
and it implies that the second cumulant in frequency ($\tilde \kappa_2(\omega) = |f(\omega)|^2$ and $f(\omega)$ in the standard ETH notations) is only a function of the spectral correlations $\frac{R_2(\omega)}{D^2}$ and a constant function $\kappa_2(A)$ which only depends on the eigenvalues of the operator. This corresponds to the standard ETH ansatz for GUE (CUE) matrices, see e.g. Ref.~\cite{liao2022field} \\

To summarize, in basis-rotationally-invariant systems, {assuming decoupling between the average of matrix elements and of the spectrum}, the free cumulant in the frequency domain is given by the product of the free cumulant of the matrix (constant in frequency) and the spectral correlations, which are constant almost everywhere, i.e. when all frequencies $\omega_{ij}$ are much larger than mean level spacing, {(see Chapter 6 of Ref.\cite{mehta2004random})}.

\section{Factorization of non-crossing contribution in the time-domain}
\label{app_factot}
The full ETH predicts the factorization of the non-crossing contribution in time [c.f. Eq.~(3) of the main text] as well as in the frequency domain. 
In the main text, we have studied the factorization at equal times and as a function of frequency in Fig.~2 of the main text. Here we discuss the factorization in the time domain. We consider the cactus contribution appearing in the OTOC of 
Eq.~(17) of the main text, namely
\begin{equation}
    \text{cactus}(t) = \text{cactus}(t, t) = \frac{1}D \sum_{i\neq j \neq m} e^{\ii\omega_{ij} t + i \omega_{im} t} {|A_{ij}|^2| A_{im}|^2} 
\end{equation}
with $\text{cactus}(t_1, t_2)$ is defined in Eq.~(15). The factorization in Eq.~(3) of the main text implies that in the large $L$ limit one shall have
\begin{equation}
    \text{cactus}(t) \simeq [k^{\rm ETH}_2(t)]^2 + \mathcal O(1/L) \ .
\end{equation}

\begin{figure}[t]
 \hspace{-1cm}
	\includegraphics[width=.33\linewidth]{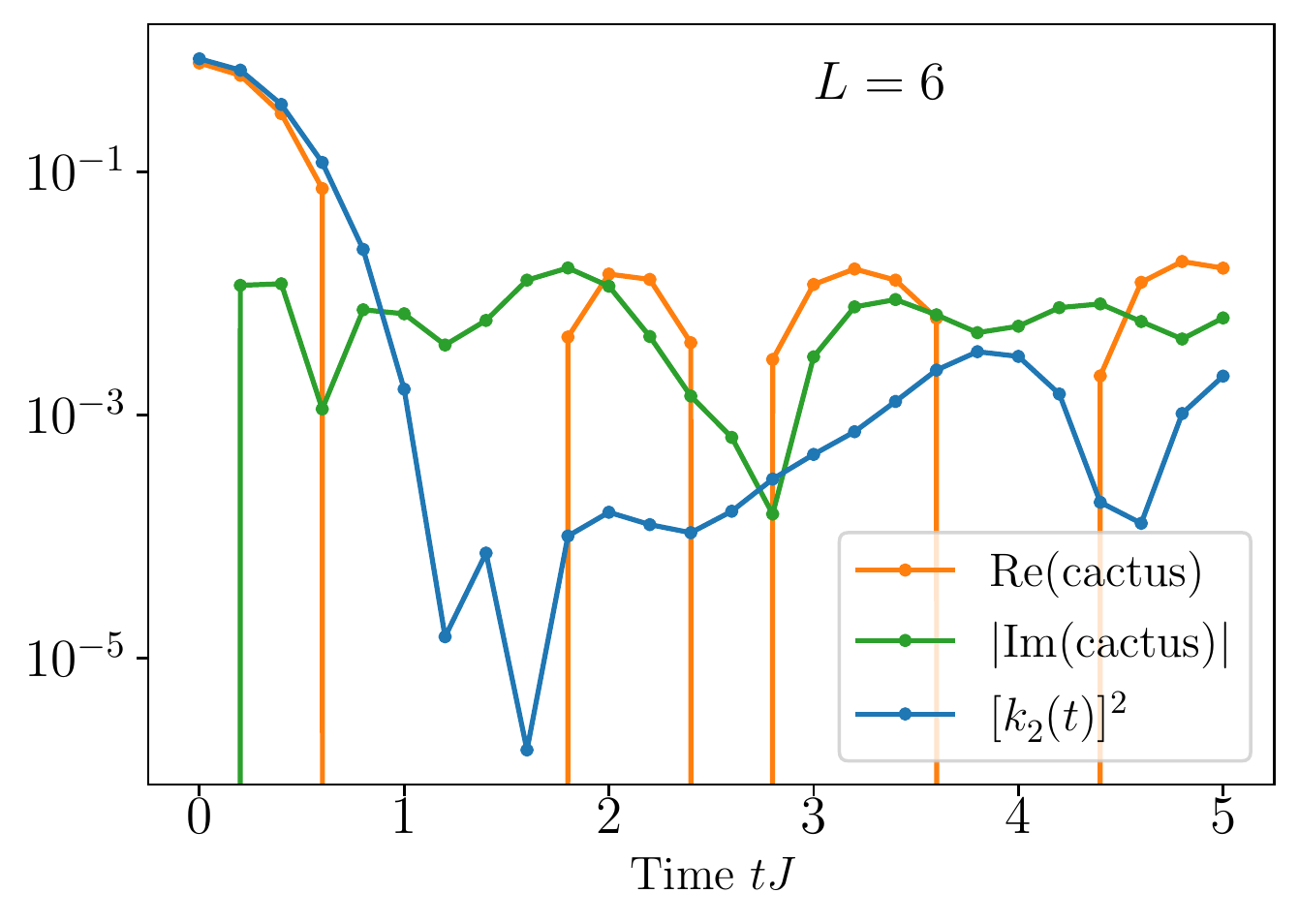}
	\includegraphics[width=.33\linewidth]{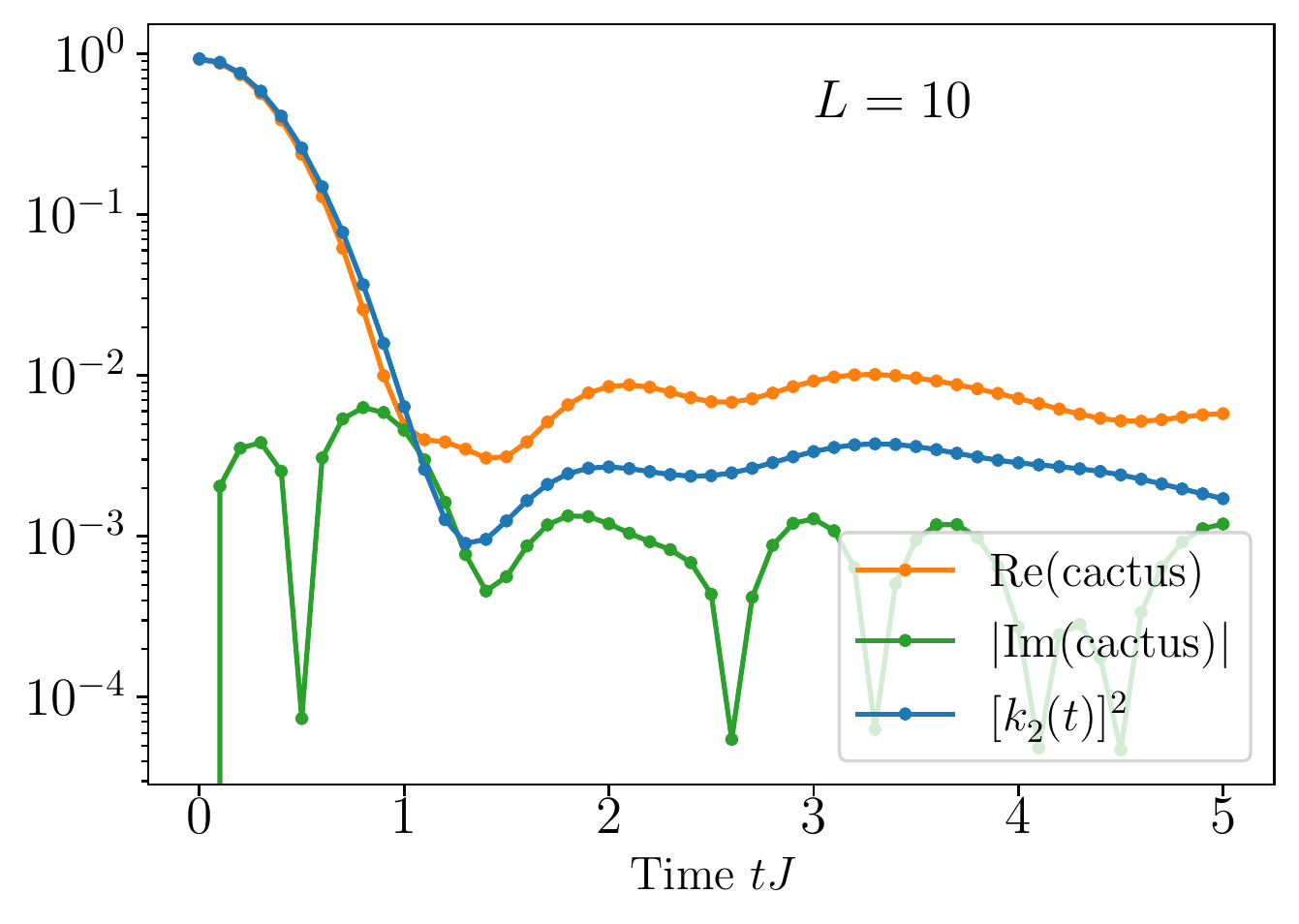}
	\includegraphics[width=.33\linewidth]{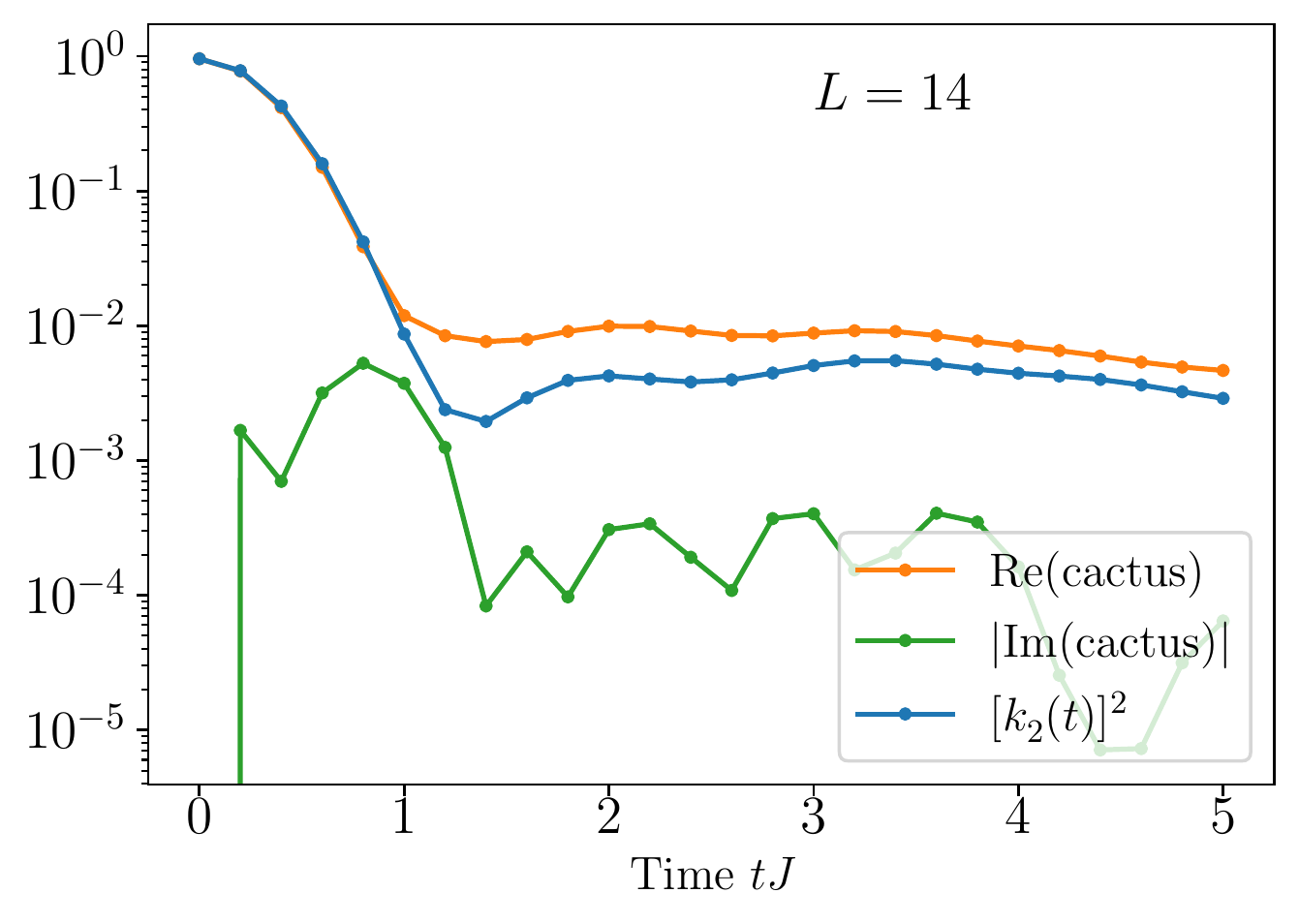}
  \hspace{-2cm}
	\includegraphics[width=.33\linewidth]{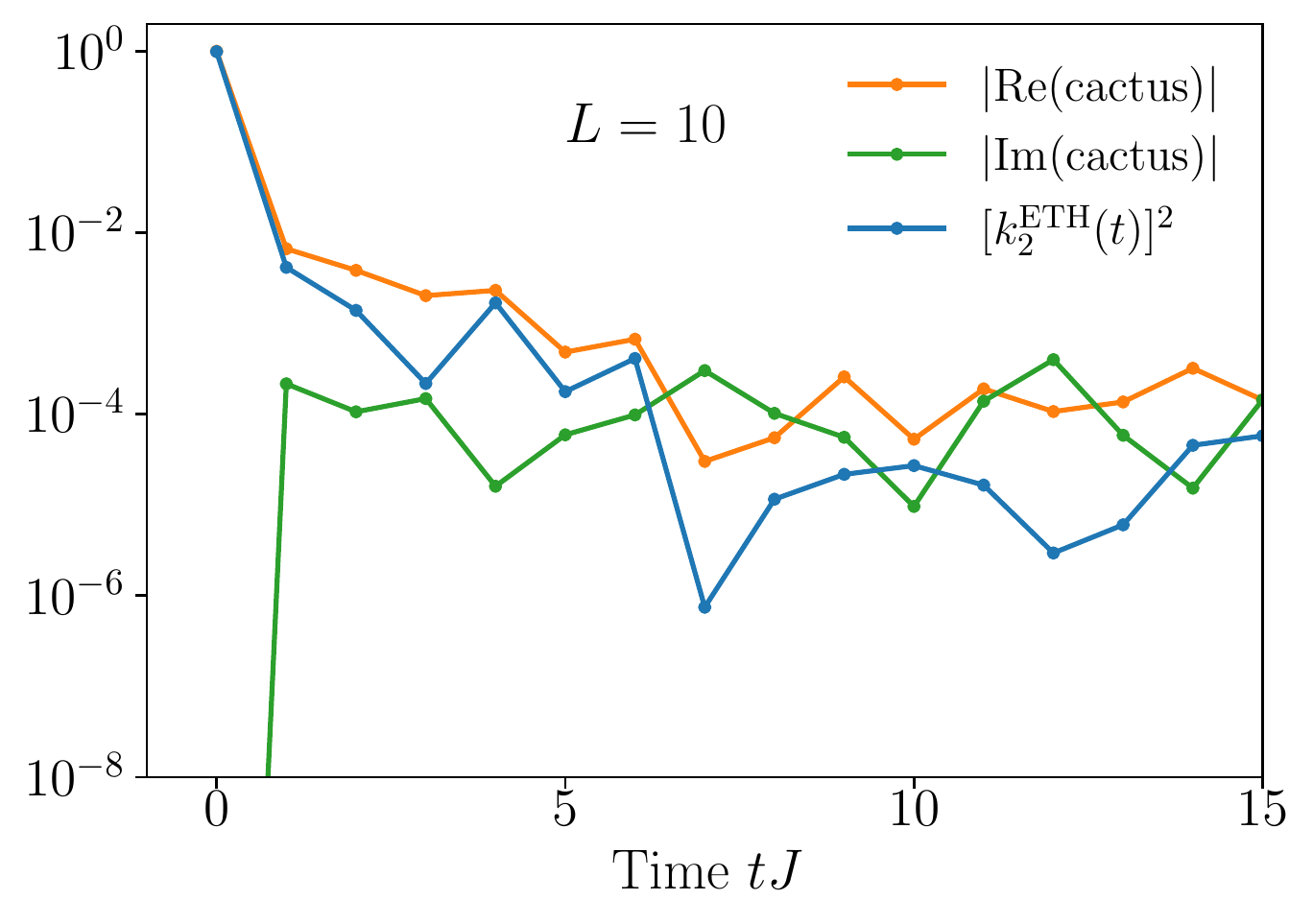}
	\includegraphics[width=.325\linewidth]{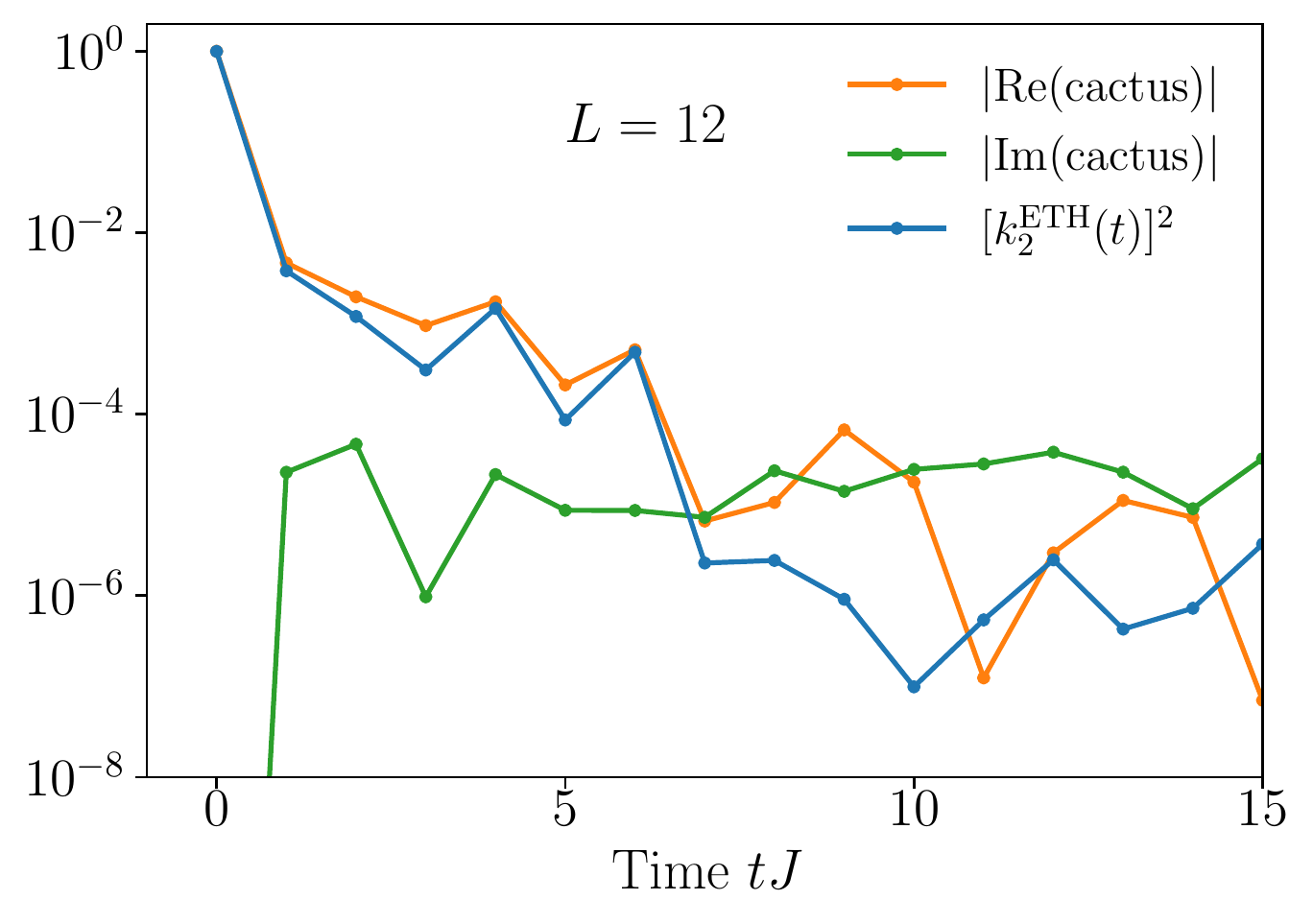}
	\includegraphics[width=.325\linewidth]{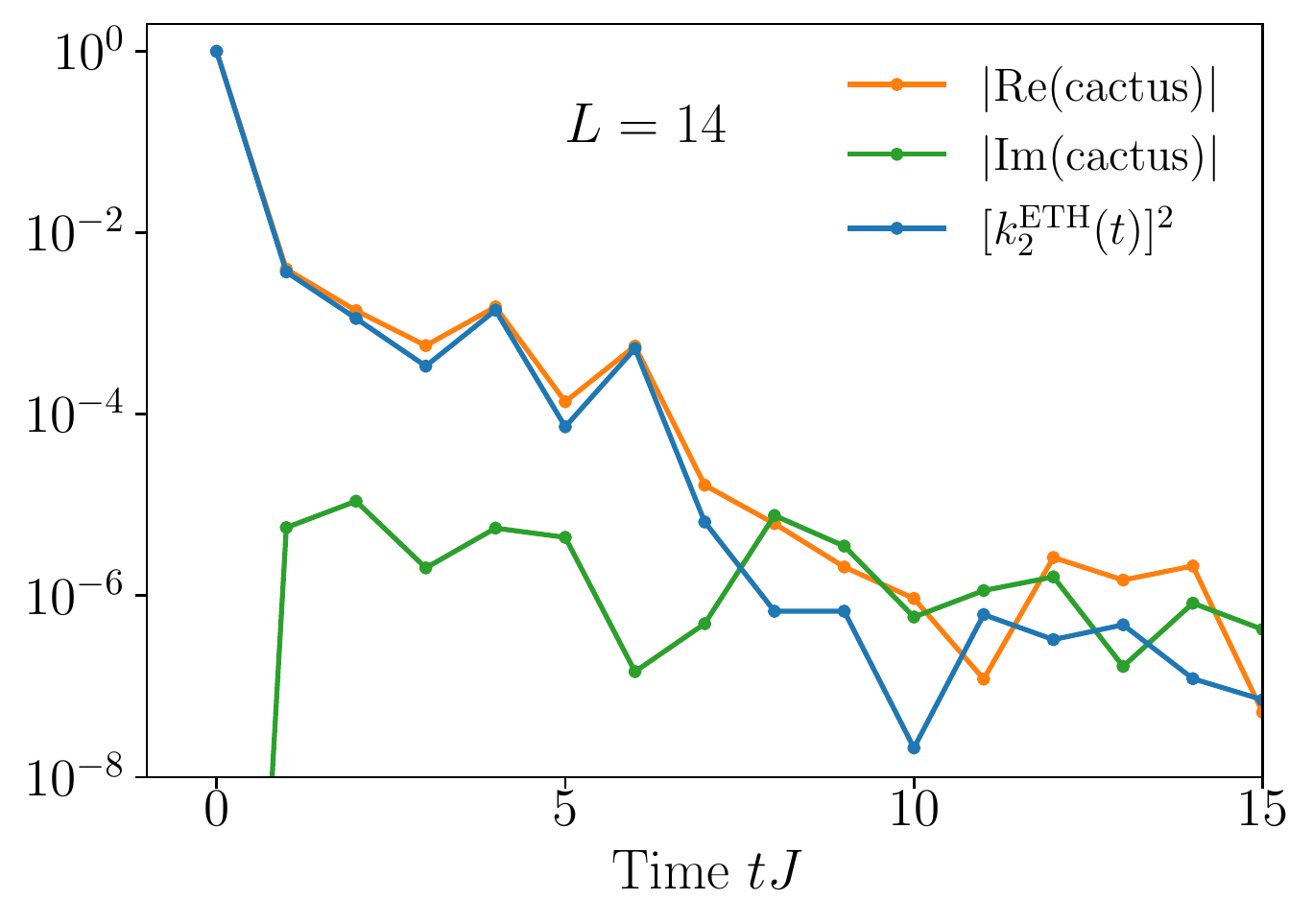}
	\label{fig:4_local}
 \caption{Factorization of the cactus diagram in the time domain. (Top row) The Ising model with system size $L=8, 12, 16$ and the collective observable. (Bottom row) Circuit with $L=10, 12, 14$ for $\sigma^z_{L/2+1}$.}
\end{figure}

This factorization is checked numerically and the data are reported in Fig.~\ref{fig:4_local} for the Hamiltonian case (top row) and the circuit (bottom row). For finite times, $\text{cactus}(t)$ develops an imaginary part, that oscillates around $t=0$ (in the figure we plot the absolute value) and it is suppressed as the system size increases. Furthermore, the agreement between $\text{Re (cactus}(t))$ and $[k^{\rm ETH}_2(t)]^2$ holds up to a time scale which diverges in the limit $L\to \infty$.

\section{Results for other local operators}
\label{app_local}
\begin{figure}[t]
\centering
\includegraphics[width=1 \linewidth]{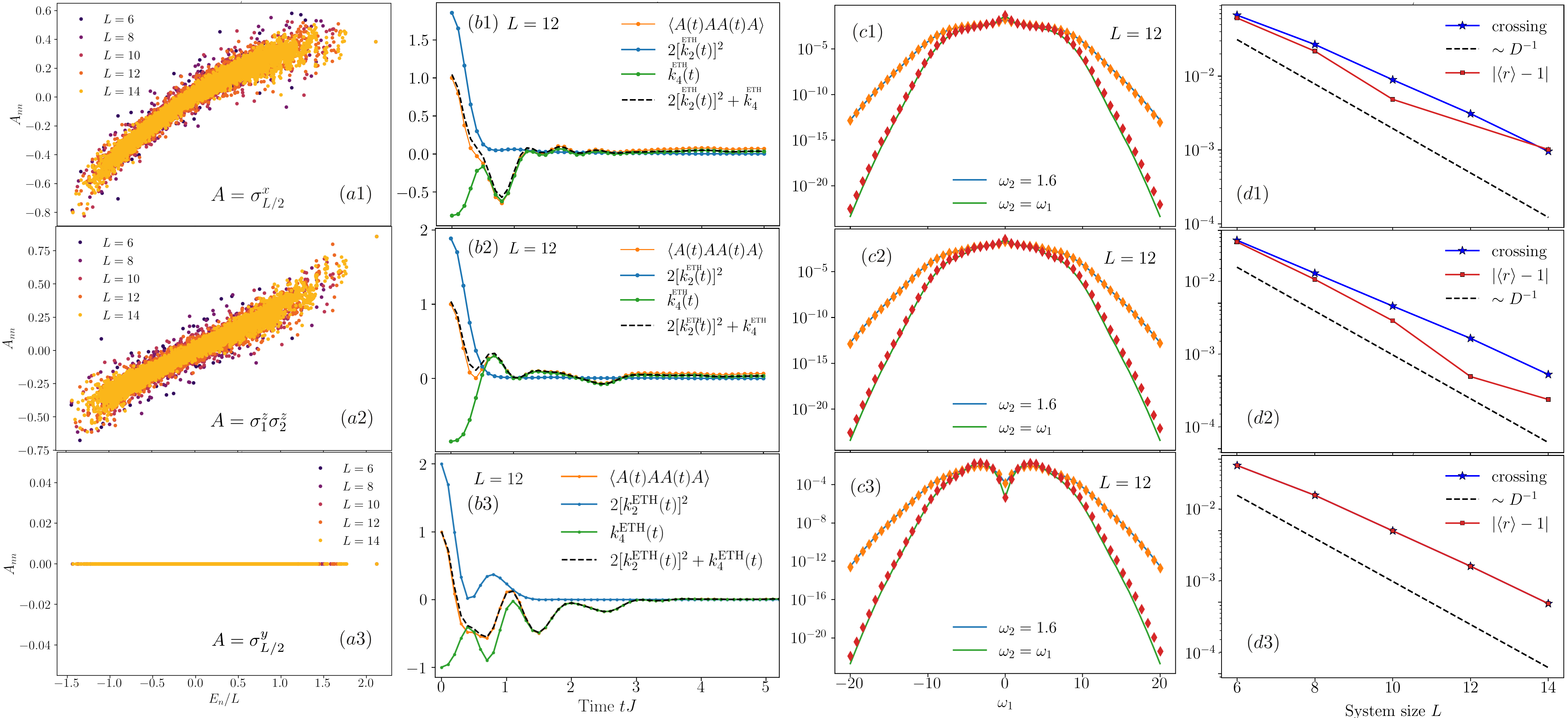}
\caption{Full ETH by free cumulants for local observables. Column 1: energy dependence of the diagonal expectation values for different system sizes $L$. Column 2: dynamics of the full OTOC compared with the free-cumulant decomposition [cf. Eq.(16) of the main text], and Column 4: factorization of the non-crossing partition in the frequency domain and Column 5: higher-order ETH conditions as a function of the system size $L$ [cf. Eq.(5-6) of the main text]. Row 1: $\hat A = \hat \sigma_1^z  \hat \sigma_2^z$, Row 2 $\hat A = \hat \sigma_{L/2}^x$ and Row 3 $\hat A = \hat \sigma_{L/2}^y $. }
\label{fig_local_ham}
\end{figure}

 In this section, we show numerically the validity the full ETH for other local observables defined on local support, such as single-site or two-site spin operators.
We exemplify our findings in the Ising model, by considering three different observables, as shown in Fig.\ref{fig_local_ham}:  $\hat A = \hat \sigma_{L/2}^x $ (Row 1)  $\hat A = \hat \sigma_1^z  \hat \sigma_2^z$ (Row 2) and $\hat A = \hat \sigma_{L/2}^y $ (Row 3). The exact diagonalization is performed in the full Hilbert, with all the matrix elements.

The first column displays the energy dependence of the diagonal expectation value $A_{nn}$ as a function of the energy density: as the system size $L$ is increased, the $A_{nn}$ becomes a smooth function of $E_n/L$. This corresponds to the well-established validity of the standard ETH, which is here displayed to emphasize the energy dependence of the considered observables. 
We note that the observable $\hat A = \hat \sigma^y_{L/2}$ (Fig.\ref{fig_local_ham}a3)  has vanishing diagonal matrix elements $A_{nn}=0$ $\forall n$, due to the property $\sigma^y_{L/2}\propto [\hat H, \hat \sigma^z_{L/2}]$.

The second column represents the dynamical moment-ETH free cumulant decomposition for the OTOC [cf. Eq.~(16) of the main text]. 
 Since $\hat A^2 = \hat A^4 = 1$, hence one has also $\langle A(0) A A(0) A\rangle = k^{\rm ETH}_2(0) = 1$. Therefore the decomposition $\langle A^4\rangle = 2 [k^{\rm ETH}_2(0)]^2+k^{\rm ETH}_4(0)$ implies that the fourth free cumulant starts from a negative value: $k^{\rm ETH}_4(0)=-1$. As in the main text, also here $k^{\rm ETH}_2(t)$ approaches zero after a short time-scale, while the characteristic OTOC dynamics is given by $k^{\rm ETH}_4(t)$. 
 \\

In the third column, we show the factorization as a function of frequency, i.e. $\text{cac}(\omega_1, \omega_2) \simeq \tilde k^{\rm ETH}_2(\omega_1) \tilde k^{\rm ETH}_2(\omega_2)$, we report the data along two directions $\omega_2=\omega_1$ and $\omega_2=1.6$. The results show a very good agreement with small deviations at large frequencies due to finite-size effects. In fact, 
such large-frequency differences are due to the entropic corrections $\propto e^{-\frac{\partial^2 S(E)}{\partial E^2}\Big|_{E=0} \omega^2/8}$. These are neglected in the standard ETH calculations \cite{srednicki1999approach, dalessio2016from} because they vanish in the thermodynamic limit for finite frequencies, see e.g. the discussion below Eq.\eqref{eq_expEntro}. They are however visible for finite $L$ at large $\omega$.

In the fourth column, we report the exponential suppression of crossing partitions [cf. Eq.~(4) of the main text] and the factorization of non-crossing parts [cf. Eq.~(5) of the main text]. We show the exponential suppression of the crossing partitions and the distance from one of the equal-times ratio $|\langle r\rangle -1|$ in Eq.~(14), obtained by increasing $L$.\\
We note that in Fig.\ref{fig_local_ham}d3 and in Fig.2c the main text, one has 
\begin{align}
    \label{res_cactuAii0}
    |\langle r\rangle -1| = |\frac{ \text{cactus}(0, 0)}{(k^{\rm ETH}_2(0))^2} - 1 |=  \text{crossing}.
\end{align}
This holds for unitary observables ($\hat A^2 = 1$) with $A_{nn}=0$ ($\overline{A_{nn}}=0$), such as the ones considered in Fig.\ref{fig_local_ham}d3 (Fig.2c of the main text). 
This can be seen by computing
\begin{align}
    \text{cactus}(0, 0) & = \frac 1D \sum_{i \neq j \neq k} |A_{ij}|^2 |A_{ik}|^2\pm \frac 1D \sum_{i \neq j}|A_{ij}|^4
    \\
    & = \frac 1D \sum_{i \neq j\, i \neq k} |A_{ij}|^2 |A_{ik}|^2
    - \text{``crossing''} \label{S17}
    \\
    & = \frac 1D \sum_i \langle i|\hat A \sum_k |k\rangle \langle k| \hat A |i \rangle \, \langle i|\hat A \sum_j |j\rangle \langle j| \hat A |i \rangle - \text{``crossing'} \label{S18}
    \\
    & = \frac 1D \sum_i \langle i|\hat A^2|i \rangle \, \langle i|\hat A^2 |i \rangle - \text{``crossing' }
    \\
    & = \frac 1D \sum_i |\langle i |i\rangle|^2 - \text{``crossing' } = 1-\text{``crossing'} \ ,
\end{align}
where in the first line with $\pm$ indicates that we add and subtract the same quantity ``{crossing}''$=\frac 1D \sum_{i \neq j}|A_{ij}|^4$ (defined in Eq.~(2) of the main text) and from Eq.~\eqref{S17} to Eq.~\eqref{S18} we have neglected the condition $i\neq j$ and $i \neq k$ since we consider here observables which obey $A_{nn}=0$ (or $\overline{A_{nn}}=0$). Thus, since we have also $k^{\rm ETH}_2(0)=1$ one immediately finds Eq.\eqref{res_cactuAii0}.

\section{Results at finite temperature}
\label{sec_finiteT}
\begin{figure*}[h]
\centering
\includegraphics[width=.44\textwidth]{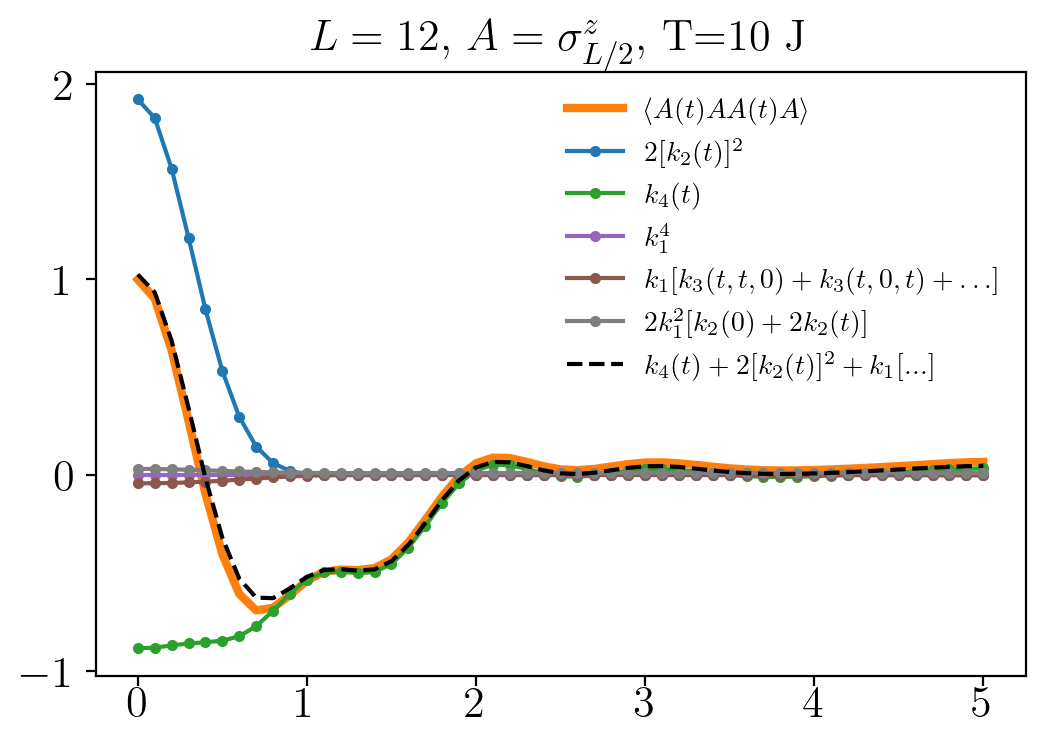}
\includegraphics[width=.44\textwidth]{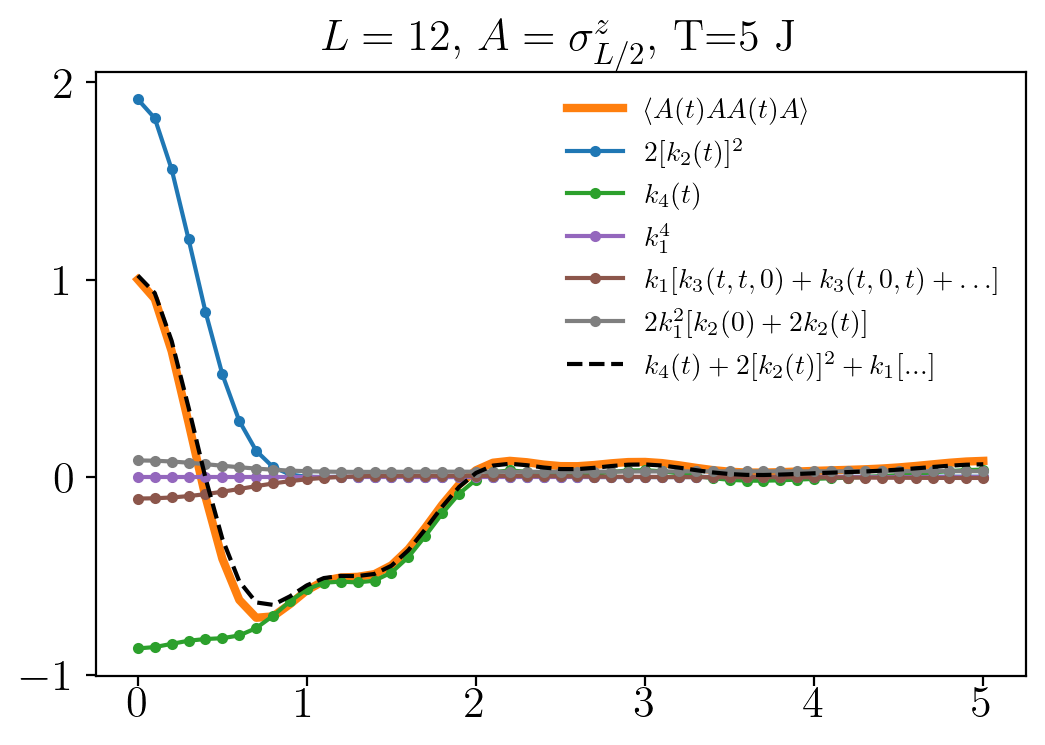}
\includegraphics[width=.44\textwidth]{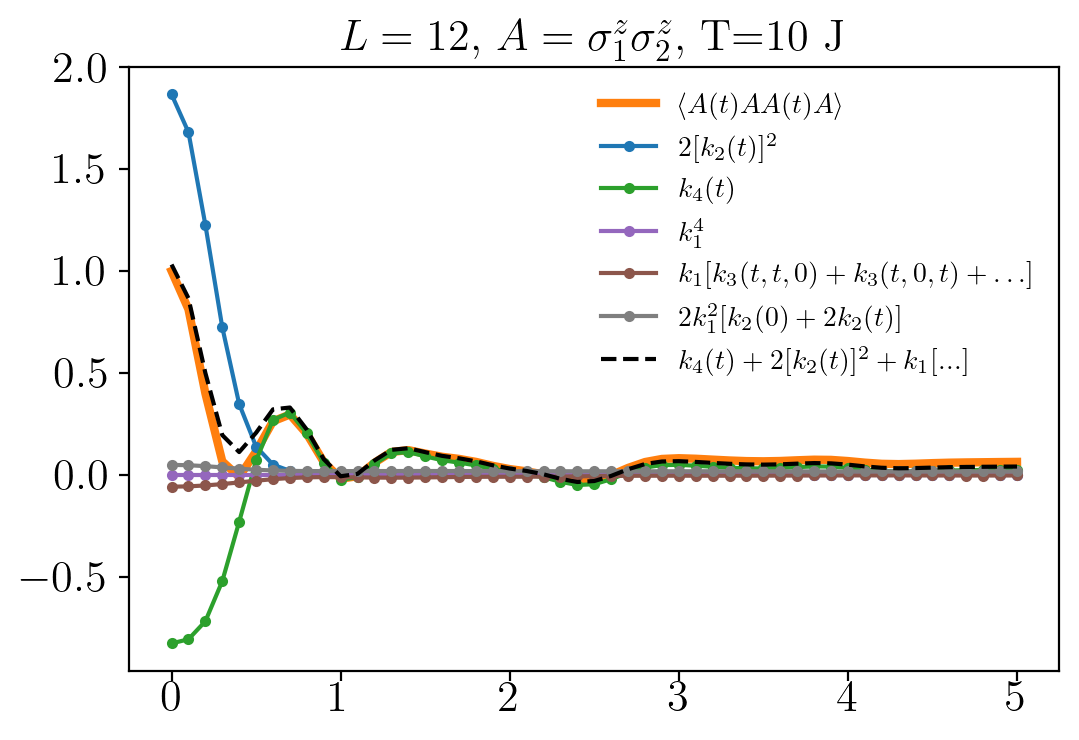}
\includegraphics[width=.44\textwidth]{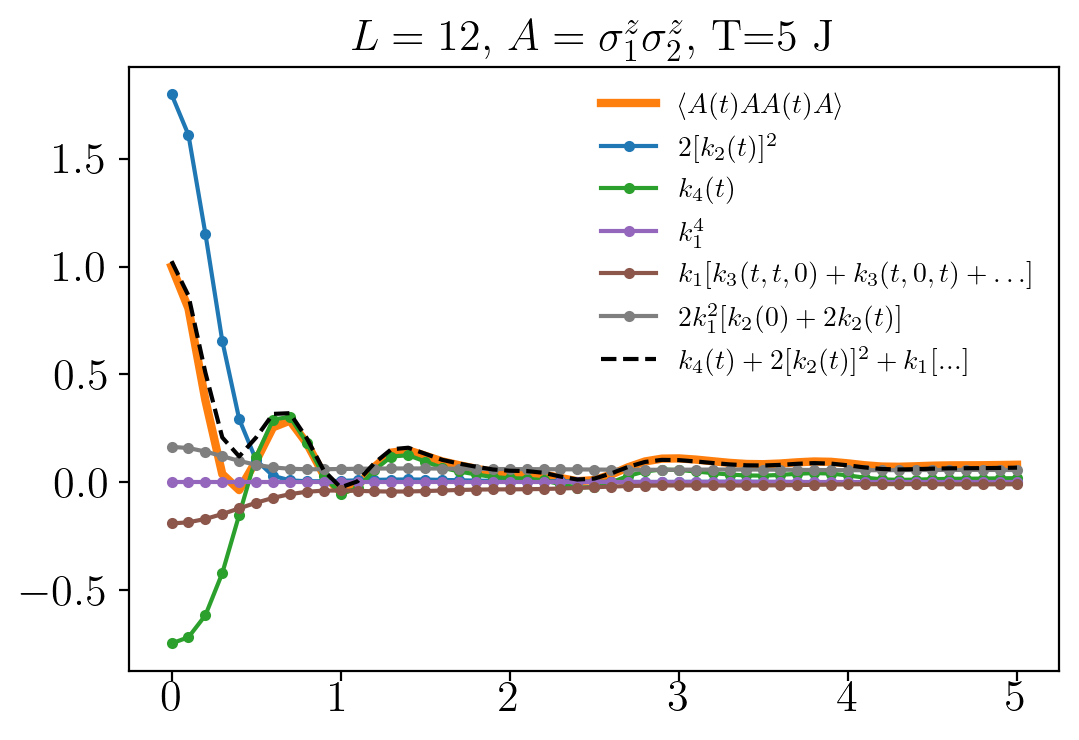}
\caption{Full ETH at finite temperature via the moment-cumulant decomposition in Eq.\eqref{eq_finitebeta}. Column 1: $T=10$ and Column 2: $T=5$. In Row 1, we consider $\hat A = \sigma_{L/2}^z$ and Row 2 $\hat A = \hat \sigma_1^z  \hat \sigma_2^z$. The full ED result for $\braket{A(t) A A(t) A}_\beta$ (orange) is compared with the ETH prediction in Eq.\eqref{eq_finitebeta} (dashed black) where in the legend with $k_1[\dots]$, the dots indicate all the terms appearing in the square brackets of Eq.\eqref{eq_finitebeta}. Analogously,  the dots in the legend of the for the brown line indicate the other two terms in the second line of Eq.\eqref{eq_finitebeta}.
 The calculations are performed in the full Hilbert space for $L=12$. 
 }
\label{fig2}
\end{figure*}

In the main text and in the plot above, we have explored correlations at infinite temperature for which $k^{\beta=0}_1=0$. We now study finite temperatures, for which $k^{\beta}_1=\braket{A}_{\beta}\neq 0$. The ETH result for four point function is given by the full Eq.\eqref{eq_20}, which for $t_1=t_3=t$ and $t_2=t_4=0$ reads:
\begin{align}
\label{eq_finitebeta}
    \begin{split}
        \langle A(t) A A(t) A \rangle_\beta = & 
        k_4^{\rm ETH}(t, 0, t, 0) 
        + 2 [k_2^{\rm ETH}(t)]^2  
        + k^{\rm ETH}_1 \Big [
        [k^{\rm ETH}_1]^3 + 2\, k^{\rm ETH}_1 k^{\rm ETH}_2(0)
       +  4 k^{\rm ETH}_1 k^{\rm ETH}_2(t)  
         \\ & 
         k^{\rm ETH}_3(t, 0, t)  
        + k^{\rm ETH}_3(t, t, 0) 
        + k^{\rm ETH}_3(t, 0, 0) 
        + k^{\rm ETH}_3(0, t, 0) 
        \Big ]\ .
    \end{split}
\end{align}
We repeat the calculations at finite temperature and we compare with the ETH result for finite temperature in Eq.\eqref{eq_finitebeta}. We show the results for $T=5$ and $T=10$ in Fig.\ref{fig2}, where we consider the local operator $\hat A=\hat \sigma^x_{L/2}$ and $\hat A=\hat \sigma^z_{1} \hat \sigma^z_{2}$.  We find agreement with the ETH predictions also at finite temperatures for local operators. 

\section{Details on the Circuit model}
\label{sec_circ}
For completeness, we report the unitary two-qubit gates used for numerical simulations in the Circuit model here. The gates are given by
\begin{equation}
    U_1  = \left( \begin{smallmatrix}
    0.67931743- \ui 0.0721112  &    0.51877751 - \ui 0.04466678 &  0.25741725 +\ui 0.08032817 &   -0.36793521 - \ui 0.23261557 \\
  0.3619221 - \ui 0.46028992  & -0.56476001+ \ui 0.43994397 &  -0.09219379+ \ui 0.36114353 &   -0.07557832- \ui 0.00214075 \\
  0.13342533 + \ui 0.37674694 & -0.30800024 + \ui 0.1579134j &  0.62386475 - \ui 0.20799404 &   -0.20374812 + \ui 0.49646409 \\
  -0.13483809- \ui 0.11203347 & -0.30258211 + \ui 0.07080587 &   0.25709566 - \ui 0.53832377 &  -0.15624451 - \ui 0.70170847
    \end{smallmatrix} \right)
\end{equation}
and 
\begin{equation}
    U_2 = \left( \begin{smallmatrix}
    -0.32936937+\ui 0.0575935 &  -0.32742441+ \ui 0.30978811 & -0.41818243 + \ui 0.42081727 & -0.14116509-\ui 0.55958207 \\
    -0.53553904-\ui 0.54762144 & 0.13993909 - \ui 0.01096924 &  0.5221732 + \ui 0.23437906 &  0.23558662-\ui 0.10249863 \\
    -0.09194887+\ui 0.40213391&  -0.35024609 + \ui 0.37068525 &   0.14797073+ \ui 0.19071808 &  0.60159551 + \ui 0.38674017 \\
    0.02435271 - \ui  0.36159117 &  -0.33621261 + \ui 0.63561192 &  0.11961071- \ui 0.49785786 &  -0.2804114 + \ui 0.10400783
    \end{smallmatrix} \right).
\end{equation}

\end{document}